\documentclass[notitlepage]{revtex4-1}
\usepackage{amsmath}
\usepackage{bm}
\usepackage{natbib}
\usepackage{graphicx}
\usepackage{hyperref}
\usepackage{color,soul}

\begin{document}
\title{High order difference schemes using the Local Anisotropic Basis Function Method}

\author{J. R. C. King}\email{jack.king@manchester.ac.uk}\affiliation{Department of Mechanical, Aerospace and Civil Engineering, The University of Manchester, Manchester, UK}
\author{S. J. Lind}\affiliation{Department of Mechanical, Aerospace and Civil Engineering, The University of Manchester, Manchester, UK}
\author{A. Nasar}\affiliation{Department of Mechanical, Aerospace and Civil Engineering, The University of Manchester, Manchester, UK}\date{\today}


\begin{abstract}
Mesh-free methods have significant potential for simulations in complex geometries, as the time consuming process of mesh-generation is avoided. Smoothed Particle Hydrodynamics (SPH) is the most widely used mesh-free method, but suffers from a lack of consistency. High order, consistent, and local (using compact computational stencils) mesh-free methods are particularly desirable.

Here we present a novel framework for generating local high order difference operators for \emph{arbitrary} node distributions, referred to as the Local Anisotropic Basis Function Method (LABFM). Weights are constructed from linear sums of anisotropic basis functions (ABFs), chosen to ensure exact reproduction of polynomial fields up to a given order. The ABFs are based on a fundamental Radial Basis Function (RBF), and the choice of fundamental RBF has small effect on accuracy, but influences stability. LABFM is able to generate high order difference operators with compact computational stencils ($4^{th}$ order with $\mathcal{N}\approx{25}$ nodes, $8^{th}$ order with $\mathcal{N}\approx{60}$ nodes in two dimensions). At domain boundaries (with incomplete support) LABFM automatically provides one-sided differences of the same order as the internal scheme, up to $4^{th}$ order. We use the method to solve elliptic, parabolic and mixed hyperbolic-parabolic partial differential equations (PDEs), showing up to $8^{th}$ order convergence. The inclusion of hyperviscosity is straightforward, and can effectively provide stability when solving hyperbolic problems.

LABFM is a promising new mesh-free method for the numerical solution of PDEs in complex geometries. The method is highly scalable, and for Eulerian schemes, the computational efficiency is competitive with RBF-FD for a given accuracy. A particularly attractive feature is that in the low order limit, LABFM collapses to Smoothed Particle Hydrodynamics (SPH), and there is potential for Arbitrary Lagrangian-Eulerian schemes with natural adaptivity of resolution and accuracy.

\end{abstract}

\maketitle

\section{Introduction\label{intro}}

The numerical solution of partial differential equations (PDEs) is key to many branches of science, and the calculation of the spatial derivatives of a field, based on the knowledge of the value of that field at a discrete set of points, is key. More accurate approximations lead to more accurate results. Approximations which converge at high order are attractive, yielding equivalent accuracy at lower resolution, and hence lower computational cost. Mesh-based methods have long dominated numerical simulations, and for simple problem geometries, can be made extremely accurate (e.g. tenth order finite differences (FD)~\cite{lele_1992}, or exponentially convergent spectral methods~\cite{gottlieb_1977}). For complex geometries, mesh-based methods suffer two main drawbacks. Firstly, the symmetries which enable efficient and highly accurate approximations no longer apply. Global spectral methods are no longer possible, and high order finite differences are only possible where an orthogonal mesh can be fitted to the problem. Secondly, the process of mesh generation is complex and time-consuming, and accuracy can become highly dependent on the quality of the mesh. Mesh-free methods have significant potential, as discretisation of the domain becomes extremely simple, and automation of the process to create a node distribution with the desired properties (e.g. fitting the boundaries and satisfying resolution criteria) is relatively straightforward. There is no need to generate and store information on inter-node connectivity. In the context of computational fluid dynamics, mesh-free methods also have the advantage that topological changes in the solution (i.e. wave overtopping) which would lead to singularities in mesh based methods, may be handled easily, provided the method is based in an appropriate frame of reference. A number of mesh-free methods have gained traction in recent decades, and for a broad overview of mesh-free methods, we refer the reader to~\cite{li_review_2002,sahil_2018}. Below we provide a brief review of the methods most relevant to the present work. In the following, we use $h$ to denote the characteristic length scale of a computational stencil, $\mathcal{N}$ to denote the typical number of nodes or particles in a stencil, and $N$ to denote the total number of computational nodes.

The Generalized Finite Difference Method (GFDM)~\cite{jensen_1972} is a mesh-free extension of the finite difference method, for arbitrary node distributions. A linear system is solved for each node to obtain a moving least squares (MLS) approximation of a function and its derivatives, based on a compact stencil, from which weights for a local discrete gradient operator are calculated. The condition of the linear systems has a dependence on the node distribution~\cite{perrone_1975}, and generally the method is applied only up to second order~\cite{benito_2007,gavete_2017}. Recently~\citet{trask_2017} has developed a generalised moving least squares (GMLS) method, in which a polynomial reconstruction is used to obtain high order approximate derivatives. The method has been extended to include variable resolution, and applied to Stokes flow in complex geometries~\cite{trask_2018}, demonstrating fourth order convergence.

Smoothed Particle Hydrodynamics is a meshfree method, originally developed for astrophysical simulations~\cite{gingold_1977,lucy_1977}, and adapted with substantial success to simulations of incompressible flows, free surface flows, fluid structure interactions, and solid mechanics~\cite{monaghan_2012}. Fluid properties and their spatial derivatives are approximated based on weighted sums of the properties at neighbouring particles, with weights determined from a smoothing kernel, and its derivatives. In~\cite{gingold_1977} smoothing kernels had a Gaussian form, although in modern SPH they are typically polynomials with compact support~\cite{monaghan_2005}. The continuous form of SPH has a theoretical convergence rate of $2$ (in $h$)~\cite{quinlan}, provided $\mathcal{N}$ is held constant (for typical smoothing kernels), but in practice this drops to typically $\sim{1}$ due to particle disorder~\cite{lind_2012}. In fact, a simple error analysis of discrete SPH shows that derivative approximations are truly zero order in $h$ (and Laplacians even divergent, depending on the formulation), although there is generally a range of resolutions and particle distributions for which convergence can be demonstrated~\cite{quinlan}. The convergence rate of SPH is dependent on $\mathcal{N}$, and this dependence is influenced by the choice of smoothing kernel. For the widely used Wendland C2 kernel (e.g.~\cite{dehnen_aly}), the method approaches the theoretical convergence rate (in $h$) of $2$ with $\mathcal{N}^{-5/2}$ (i.e. the total error is $\mathcal{O}\left(h^{2}\right)+\mathcal{O}\left(\mathcal{N}^{-5/2}\right)$). A number of corrections to SPH have been devised, including by~\citet{bonet_lok}, whose relatively cheap correction (which requires inversion of a square matrix of rank $\mu$, where $\mu$ is the number of spatial dimensions) provides first order consistency. 

Higher order consistency corrections for SPH have been devised, generally based on eliminating errors via the solution of $N$ small linear systems, and many of these corrections have become separate methods in their own right. The Reproducing Kernel Particle Method (RKPM)~\cite{liu_1995,liu_1995b}, and corrective SPH (CSPH)~\cite{li_review_2002} improve the accuracy of SPH approximations by introducing a polynomial factor in the kernel function chosen to ensure consistency. RKPM has gained significant popularity in solid mechanics~\cite{li_review_2002}, and is most commonly employed in a Galerkin formulation. \citet{chen_1999} and~\citet{chen_2000} developed the Corrective Smoothed Particle Method (CSPM), in which a Taylor series expansion of the unknown function was used to derive first order consistent interpolations and derivatives approximations. \citet{zhang_2004} and~\citet{liu_2005} independently developed a correction to SPH interpolation, termed the Finite Particle Method (FPM), in which higher order was achieved by including second and higher order kernel derivatives, and solving a linear system to obtain approximations to a function and its gradient. Results were presented for one- and two-dimensional problems, up to $3^{rd}$ order, although the consistency of the approximation deteriorated in the vicinity of boundaries. \citet{liu_2006} proposed a variation of CSPM, where a function and all its derivatives (to a given order) are calculated simultaneously. This method provides consistency at boundaries, where computational stencils lack complete support. ~\citet{asprone_2010} developed a modified FPM, with the kernel modified by multipication with a polynomial to eliminate certain moments, and in~\cite{asprone_2011} extended the idea by replacing the kernel and its derivatives with a series of monomials, demonstrating second order convergence in two dimensions. Building on~\cite{liu_2006},~\citet{sibilla_2015} presented a second order correction, involving the solution of a rank $5$ (in two dimensions) linear system, to eliminate low order errors. The method of~\citet{sibilla_2015} provides a consistency correction for both gradient and Laplacian operators, yielding convergence rates which do not deteriorate at very fine resolutions: the order collapses only when machine precision errors start to dominate. Recent work by~\citet{lind_2016} has shown that higher than second order in $h$ may be achieved in SPH by modifying the SPH kernel to eliminate certain moments (in the continuous form). Convergence up to sixth order has been demonstrated for Eulerian (with the particles fixed in space) schemes with these high order kernels, and the approach has been succesfully extended to semi-Eulerian schemes~\cite{fourtakas_2018}. Whilst this approach is promising, zero-order (in $h$) error terms remain for disordered particle distributions, although in practice there exists a range of resolutions for which high order convergence may be observed. This range depends on how well the discrete sums over neighbour particles approximate continuous integrals. Extending the range of resolutions for which high order is observed may be achieved either by enforcing a certain degree of uniformity and isotropy on the particle distribution, or by use of a modified Gaussian kernel combined with an extremely large computational stencil. With $\mathcal{N}\approx450$ nodes in two dimensions, corresponding to $12$ particles across the support radius, the discretisation error is $<10^{-16}$ for Gaussian kernels~\cite{trefethen_2014}. In practice, the range of resolutions for which high order is observed is dependent on (for a $4^{th}$ order kernel) the ratio of the third and fifth discrete moments of the kernel, and the solution itself. The computational cost per particle per time-step is at best $\mathcal{O}\left(\mathcal{N}\right)$ FLOPs (FLoating Point Operations), and so whilst the latter approach can yield convergence to machine precision, it increases the cost over compactly supported kernels, which typically require $\mathcal{N}\le50$ in two dimensions.

A featured shared by GFDM, GMLS and the above mentioned SPH-derived methods is that interpolation involves summation over all $\mathcal{N}$ nodes in the computational stencil of a local kernel function. Radial Basis Function (RBF) methods, which originated in the field of cartography as an interpolation method for multivariate scattered node data~\cite{hardy_1971}, and were first applied to the solution of PDEs in~\cite{kansa_1990a,kansa_1990b}, provide a different approach. An RBF is located at each computational node, with each RBF weighted such that the sum of \emph{all $N$} RBFs provides exact interpolation at the nodes. Early RBF methods were global, with weights obtained by the solution of a dense global linear system (of size $N\times{N}$). Global RBF methods can achieve exponential convergence and spectral accuracy~\cite{fornberg_2015}, although the cost of the global approach is significant, at $\mathcal{O}\left(N^{2}\right)$ FLOPs per time-step~\cite{bollig_2012}. The linear systems in global RBF methods also suffer from poor conditioning. To overcome these shortfalls a local version of the RBF method - the RBF finite difference (RBF-FD) method - was developed independently by~\citet{shu_2003,cecil_2004,wright_2003}, where each node only interacts with the nearest $\mathcal{N}$ nodes (and generally $\mathcal{N}\ll{N}$). In RBF-FD, the approximations of a function and its derivative are constructed from a set of weights obtained by solving a small linear system at every node. The method is extremely fast, scalable and accurate, although the exponential convergence and spectral accuracy of the global method is lost. Provided $\mathcal{N}\ll{N}$, the cost is $\mathcal{O}\left(N\right)$ FLOPs per time-step, with the costs of assembling and solving the local linear systems relegated to the preprocessing stage. In RBF-FD, the order of convergence is dependent on the stencil size, and (in two-dimensions) approximately $\sqrt{\mathcal{N}}$. Recent work has shown that the addition of polynomials to the local linear systems in RBF-FD can control the order of convergence, and improve the conditioning of the linear systems~\cite{flyer_2016,flyer_2016_ns}. A good review of both global RBF and RBF-FD methods is given by~\citet{fornberg_2015}. 

In this paper we present a novel technique for calculating high order difference approximations, which we term the Local Anisotropic Basis Function method (LABFM). Local discrete operators approximating the spatial derivatives are constructed from a linear combination of anisotropic basis functions (ABFs). The weights for the linear sum of ABFs are determined to enforce specific error terms to be zero, ensuring exact reproduction of polynomial fields up to the desired order, on arbitrary node distributions. In this work we demonstrate up to $8^{th}$ order convergence on disordered node distributions. 

The remainder of the paper is set out as follows. In Section~\ref{error} we introduce our notation, a general discrete operator, and an expresion for the error therein. In Section~\ref{labf} we present our method for calculating weights for high order difference operators, based on a series of Anisotropic Basis Functions (ABFs), derived from a fundamental RBF. In Section~\ref{conv} we present numerical results using our method focusing on the accuracy and stability. In Section~\ref{pde} we demonstrate the ability of our method to solve a range of prototype PDEs. Section~\ref{conc} is a summary of conclusions. In Appendix~\ref{rbf_details} we provide details of the ABFs used in this work.

\section{A general discrete operator}\label{error}

Consider a set of $N$ nodes distributed in a domain $\Omega$, with $\bm{r}_{i}=\left(x_{i},y_{i}\right)$ the position vector of node $i$. Throughout this paper we consider two dimensional problems for simplicity of exposition, although the method could be easily extended to higher dimensional problems (at computational cost). \emph{Note that the analysis in this section is for arbitrary node distributions.} A characteristic length scale of the distribution is $\delta{r}$, where typically $N\delta{r}^{2}=V_{\Omega}$, with $V_{\Omega}$ the volume of the domain. The value of some field $\left(\cdot\right)$ at node $i$ is $\left(\cdot\right)_{i}$. When calculating derivatives, we use stencils with finite size with characteristic width $h$ (in the present work, all stencils have a radius of $2h$). For any node $i$, we refer to all nodes $j$ within the stencil of $i$ as neighbours of $i$. $\mathcal{N}_{i}$ is the number of neighbours of $i$. Sums over $j$ (e.g. $\sum_{j}$) are over all $j\in\mathcal{N}_{i}$. In this paper, we consider the case where $h/\delta{r}$ is uniform and constant, and hence also $\mathcal{N}_{i}$) is approximately uniform and constant. We use $H$ as a characteristic length scale of the domain, or of some scalar or vector field, and then $h/H$ is a non-dimensional measure of the effective resolution. We denote $\left(\cdot\right)_{ij}=\left(\cdot\right)_{i}-\left(\cdot\right)_{j}=-\left(\cdot\right)_{ji}$, and where the subscript $ji$ appears after a function, the difference is applied to the function arguments. We define a vector of monomials
\begin{equation}\bm{X}=\begin{bmatrix}x,&y,&\frac{x^{2}}{2},&xy,&\frac{y^{2}}{2},&\frac{x^{3}}{6},&\frac{x^{2}y}{2},&\frac{xy^{2}}{2},&\frac{y^{3}}{6},&\frac{x^{4}}{24},\dots\end{bmatrix}^{T},\end{equation}
noting that $\bm{X}_{ji}=\left[x_{ji},y_{ji},x^{2}_{ji}/2,\dots\right]^{T}$ (i.e. $\bm{X}_{ji}\ne\bm{X}_{j}-\bm{X}_{i}$). We also define a vector operator of partial derivatives as
\begin{equation}\bm{D}\left(\cdot\right)=\begin{bmatrix}\frac{\partial\left(\cdot\right)}{\partial{x}},&\frac{\partial\left(\cdot\right)}{\partial{y}},&\frac{\partial^{2}\left(\cdot\right)}{\partial{x}^{2}},&\frac{\partial^{2}\left(\cdot\right)}{\partial{x}\partial{y}},&\frac{\partial^{2}\left(\cdot\right)}{\partial{y}^{2}},&\frac{\partial^{3}\left(\cdot\right)}{\partial{x}^{3}},&\frac{\partial^{3}\left(\cdot\right)}{\partial{x}^{2}\partial{y}},&\frac{\partial^{3}\left(\cdot\right)}{\partial{x}\partial{y}^{2}},&\frac{\partial^{3}\left(\cdot\right)}{\partial{y}^{3}},&\frac{\partial^{4}\left(\cdot\right)}{\partial{x}^{4}},\dots\end{bmatrix}^{T}.\end{equation}
With this notation, the multi-dimensional Taylor expansion of $\left(\cdot\right)$ about $i$ may be expressed concisely as
\begin{equation}\left(\cdot\right)_{j}=\left(\cdot\right)_{i}+\bm{X}_{ji}\cdot\left.\bm{D}\left(\cdot\right)\right\rvert_{i},\label{eq:tsc}\end{equation}
in which the subscript $i$ in the final term on the RHS indicates $\bm{D}$ is evaluated at $\bm{r}_{i}$. We define a general discrete operator, operating on node $i$:
\begin{equation}L^{d}_{i}\left(\cdot\right)=\displaystyle\sum_{j}\left(\cdot\right)_{ji}w^{d}_{ji}\label{eq:general_do}\end{equation}
where $d$ identifies the specific partial derivative(s) being approximated, and $w^{d}_{ji}$ are the set of weights that, at this stage we require to be functions of $\bm{r}_{ji}$ (i.e. $w^{d}_{ji}=w^{d}\left(\bm{r}_{ji}\right)$) and differentiable. Equation~\eqref{eq:general_do} approximates $\left.\bm{C^{d}}\cdot\bm{D}\left(\cdot\right)\right\rvert_{i}$ where $\bm{C^{d}}$ is a vector pointing to the appropriate derivatives for a given $d$. Usually we are interested in estimating the gradients and Laplacian of a function, for which cases we set $d=x$, $d=y$, and $d=L$, giving
\begin{equation}\bm{C^{d}}=\begin{cases}\begin{bmatrix}1,&0,&0,&0,&0,&0,&0,&0\dots\end{bmatrix}^{T}&\text{if }d=x\\
\begin{bmatrix}0,&1,&0,&0,&0,&0,&0,&0\dots\end{bmatrix}^{T}&\text{if }d=y\\\begin{bmatrix}0,&0,&1,&0,&1,&0,&0,&0\dots\end{bmatrix}^{T}&\text{if }d=L.\end{cases}\end{equation}
The choice to use $\left(\cdot\right)_{ji}$ in~\eqref{eq:general_do} (as opposed to $\left(\cdot\right)_{j}$) allows inter-node weight functions with a singularity at $w^{d}_{ii}$. We analyse the error in $L^{d}_{i}$ by substituting the Taylor expansion~\eqref{eq:tsc} into~\eqref{eq:general_do}, obtaining
\begin{equation}L^{d}_{i}\left(\cdot\right)=\displaystyle\sum_{j}\left.\bm{X}_{ji}\cdot\bm{D}\left(\cdot\right)\right\rvert_{i}w^{d}_{ji},\label{eq:error_short}\end{equation}
which when expanded is
\begin{multline}L^{d}_{i}\left(\cdot\right)=\left.\frac{\partial\left(\cdot\right)}{\partial{x}}\right\rvert_{i}\displaystyle\sum_{j}x_{ji}w^{d}_{ji}+\left.\frac{\partial\left(\cdot\right)}{\partial{y}}\right\rvert_{i}\displaystyle\sum_{j}y_{ji}w^{d}_{ji}+\left.\frac{\partial^{2}\left(\cdot\right)}{\partial{x}^{2}}\right\rvert_{i}\displaystyle\sum_{j}\frac{x_{ji}^{2}}{2}w^{d}_{ji}\\+\left.\frac{\partial^{2}\left(\cdot\right)}{\partial{x}\partial{y}}\right\rvert_{i}\displaystyle\sum_{j}x_{ji}y_{ji}w^{d}_{ji}+\left.\frac{\partial^{2}\left(\cdot\right)}{\partial{y}^{2}}\right\rvert_{i}\displaystyle\sum_{j}\frac{y_{ji}^{2}}{2}w^{d}_{ji}+\left.\frac{\partial^{3}\left(\cdot\right)}{\partial{x}^{3}}\right\rvert_{i}\displaystyle\sum_{j}\frac{x_{ji}^{3}}{6}w^{d}_{ji}+\left.\frac{\partial^{3}\left(\cdot\right)}{\partial{x}^{2}\partial{y}}\right\rvert_{i}\displaystyle\sum_{j}\frac{x_{ji}^{2}y_{ji}}{2}w^{d}_{ji}\\+\left.\frac{\partial^{3}\left(\cdot\right)}{\partial{x}\partial{y}^{2}}\right\rvert_{i}\displaystyle\sum_{j}\frac{x_{ji}y_{ji}^{2}}{2}w^{d}_{ji}+\left.\frac{\partial^{3}\left(\cdot\right)}{\partial{y}^{3}}\right\rvert_{i}\displaystyle\sum_{j}\frac{y_{ji}^{3}}{6}w^{d}_{ji}+\left.\frac{\partial^{4}\left(\cdot\right)}{\partial{x}^{4}}\right\rvert_{i}\displaystyle\sum_{j}\frac{x_{ji}^{4}}{24}w^{d}_{ji}+\dots.\label{eq:error_long}\end{multline}
We refer to the sums on the RHS as the moments of $w^{d}$. The $m$-th moments are denoted $B^{d,m}$, where $m$ is the total order of the monomials within the sum. There are two first moments, three second moments, four third moments, and so on. The vector of moments is defined
\begin{equation}\bm{B^{d}}_{i}=\displaystyle\sum_{j}\bm{X}_{ji}w^{d}_{ji}.\label{eq:moments}\end{equation}
To ensure that $L^{d}$ is non-divergent, we require the terms in~\eqref{eq:error_long} of order $k$ to scale with $h^{k-l}$, where $l$ is the order of the derivative being approximated. For first derivatives, this can be achieved by ensuring that $w^{d}$ scales with $1/h$. If this condition is satisfied, $B^{d,m}\propto{h}^{m-1}$, and $L^{d}$ is zero-order in $h$ for first derivatives. The equivalent condition for second derivatives requires that $w^{d}$ scales with $1/h^{2}$, and that the first moments of $w^{d}$ are zero. Finite difference schemes satisfy these conditions, as do first derivative approximations in SPH, which are obtained by setting (e.g. for $\partial\left(\cdot\right)/\partial{x}$)
\begin{equation}w^{x}_{ji}=\frac{\partial{W}_{ij}}{\partial{x}}\delta{V},\end{equation}
where $W$ is the SPH smoothing kernel, and $\delta{V}$ is a nominal node volume. The use of a normalized SPH kernel (i.e. $\int_{V}WdV=1$, where $V$ is the support volume of the kernel) ensures that the approximation is zero order. A zero-order (and non-zero error) operator is not particularly useful. By manipulation of the specific form of $w^{d}$, the magnitudes of moments corresponding to the desired derivative may be brought closer to unity, and the magnitude of other moments may be reduced towards zero, such that the method shows higher (than zero) order convergence over a range of resolutions. In the next section we present a method for doing this and achieving arbitrarily high order difference operators.

\section{The Local Anistropic Basis Function method}\label{labf}
\subsection{The elimination of error terms}
We now present a method of specifying $w^{d}$ such that the error in $L^{d}$ is of arbitrarily high order. We set $w^{d}$ equal to the weighted sum of a series of Anistropic Basis Functions (ABFs) $W_{ji}=W\left(\bm{r}_{ji}/h\right)$, writing
\begin{equation}w^{d}_{ji}=\bm{W}_{ji}\cdot\bm{\Psi^{d}}_{i}=W^{1}_{ji}\Psi^{d}_{i,1}+W^{2}_{ji}\Psi^{d}_{i,2}+W^{3}_{ji}\Psi^{d}_{i,3}+W^{4}_{ji}\Psi^{d}_{i,4}+\dots,\label{eq:w_abf}\end{equation}
where the vector of basis functions is $\bm{W}_{ji}=\left[W^{1}_{ji},W^{2}_{ji},W^{3}_{ji}\dots\right]^{T}$ and the vector of weights is $\bm{\Psi^{d}}_{i}=\left[\Psi^{d}_{i,1},\Psi^{d}_{i,2},\Psi^{d}_{i,3}\dots\right]^{T}$. The basis functions are anisotropic, in that they depend on $\bm{r}_{ji}$ (as opposed to the dependence on $\left\lvert\bm{r}_{ji}\right\rvert$ of RBFs). 
Substituting~\eqref{eq:w_abf} into~\eqref{eq:moments} gives
\begin{equation}\bm{B^{d}}_{i}=\displaystyle\sum_{j}\bm{X}_{ji}\bm{W}_{ji}\cdot\bm{\Psi^{d}}_{i},\end{equation}
which may be easily reformulated as a linear system
\begin{equation}\bm{M}_{i}\bm{\Psi^{d}}_{i}=\bm{B^{d}}_{i},\label{eq:lsys1}\end{equation}
in which 
\begin{equation}\bm{M}_{i}=\displaystyle\sum_{j}\bm{X}_{ji}\otimes\bm{W}_{ji}.\label{eq:disc_form}\end{equation}
We obtain the weights $\bm{\Psi^{d}}_{i}$ by specifying desired values of each moment, replacing $\bm{B}_{i}$ in~\eqref{eq:lsys1} with $\bm{C^{d}}$ and solving the resulting system
\begin{equation}\bm{M}_{i}\bm{\Psi^{d}}_{i}=\bm{C^{d}}.\label{eq:lsys}\end{equation}
Having solved~\eqref{eq:lsys} we use~\eqref{eq:w_abf} to obtain the weights $w^{d}_{ji}$. 

The above analysis imposes no constraint on the lengths of $\bm{X}$, $\bm{W}$, $\bm{C}$, and the size of $\bm{M}$, although for consistency between~\eqref{eq:disc_form} and~\eqref{eq:lsys}, we require $\bm{X}$, $\bm{W}$ and $\bm{C}$ to be of equal length, and $\bm{M}$ square. In practice they are finite, and we use only the first $p$ elements from $\bm{X}$ and $\bm{W}$ to construct and solve~\eqref{eq:lsys}. We note here the useful relation between the number of elements $p$, and the order $k$ of the $p$-th element of $\bm{X}$:
\begin{equation}p=\frac{k^{2}+3k}{2}=2,5,9,14,20,27\dots\text{ for }k=1,2,3,4,5\dots,\end{equation}
which is valid for two-dimensional problems. If the method were applied to three-dimensional problems $p$ would be the sum of all triangular numbers from $2$ to $k+1$:
\begin{equation}p=\displaystyle\sum_{m=2}^{m=k+1}m\left(m+1\right)/2=3,9,19,34,55\dots\text{ for }k=1,2,3,4,5\dots.\end{equation}
 When~\eqref{eq:lsys} is solved with $\bm{M}_{i}$ having size $p\times{p}$, exact polynomial reconstruction of order $k$ is ensured. That is to say, the leading order (in $h$) error terms in $L_{i}^{d}$ scale with $h^{k-l+1}$, where $l$ is the order of the derivative approximated by $L_{i}^{d}$. For first derivatives, $l=1$ and the leading order error scales with $h^{k}$. For second derivatives, $l=2$ and the leading order error scales with $h^{k-1}$. The matrices $\bm{M}_{i}$ and the resulting $w^{d}$ are dependent on the anisotropy of the node distribution. However, in setting $w^{d}$ via~\eqref{eq:w_abf} and the solution of~\eqref{eq:lsys}, the accuracy of $L_{i}^{d}$ is maintained, \emph{provided the node distribution adequately samples the ABFs.} We discuss this criteria further in the following sections. Throughout this work we use $k$ - the order of polynomial reconstruction - to identify the order of the scheme used.
The matrix $\bm{M}_{i}$ is the same for all operators (different $d$), and so we calculate each $\bm{M}_{i}$ once, then solve~\eqref{eq:lsys} for each $d$ to obtain the required operators. Once~\eqref{eq:lsys} has been solved, we calculate the weights $w^{d}_{ji}$ and store them. When the method is used for solving PDEs in an Eulerian framework, the bulk of the cost is at startup, the cost of applying the operators $L_{i}^{d}$ is low - the run-time efficiency is comparable with finite difference methods - and the method is highly scalable.

\subsection{Constructing appropriate ABFs\label{abf}}

To obtain an appropriate set of ABFs, we start with the requirement that each ABF in $\bm{W}_{ji}$ approximates a different partial derivative, so that if $w^{d}_{ji}=\bm{W}_{ji}\cdot\bm{C^{d}}$
\begin{equation}L^{d}_{i}\left(\cdot\right)=\displaystyle\sum_{j}\left(\cdot\right)_{ji}w^{d}_{ji}=\displaystyle\sum_{j}\left(\cdot\right)_{ji}\bm{W}_{ji}\cdot\bm{C^{d}}\approx\left.\bm{D}\left(\cdot\right)\right\rvert_{i}\cdot\bm{C^{d}},\label{eq:approx_req}\end{equation}
for any $d$. This approximation need not be very accurate. We define the inner product as
\begin{equation}\left\langle{f}\left(\bm{r}_{ji}\right),{g}\left(\bm{r}_{ji}\right)\right\rangle=\frac{1}{V_{i}}\displaystyle\int_{V_{i}}f\left(\bm{r}_{ji}\right)g\left(\bm{r}_{ji}\right)dV_{j},\end{equation}
where $f$ and $g$ are any two functions of $\bm{r}$. Note that the moments $\bm{B^{d}}_{i}$ in~\eqref{eq:moments} are constructed from the discrete form of $\left\langle\bm{X}_{ji},w^{d}_{ji}\right\rangle$. We denote the $m^{th}$ element of $\bm{X}_{ji}$ as $X_{ji}^{m}$, and the $n^{th}$ ABF in $\bm{W}_{ji}$ as $W_{ji}^{n}$. 

\begin{figure}
\includegraphics[width=\textwidth]{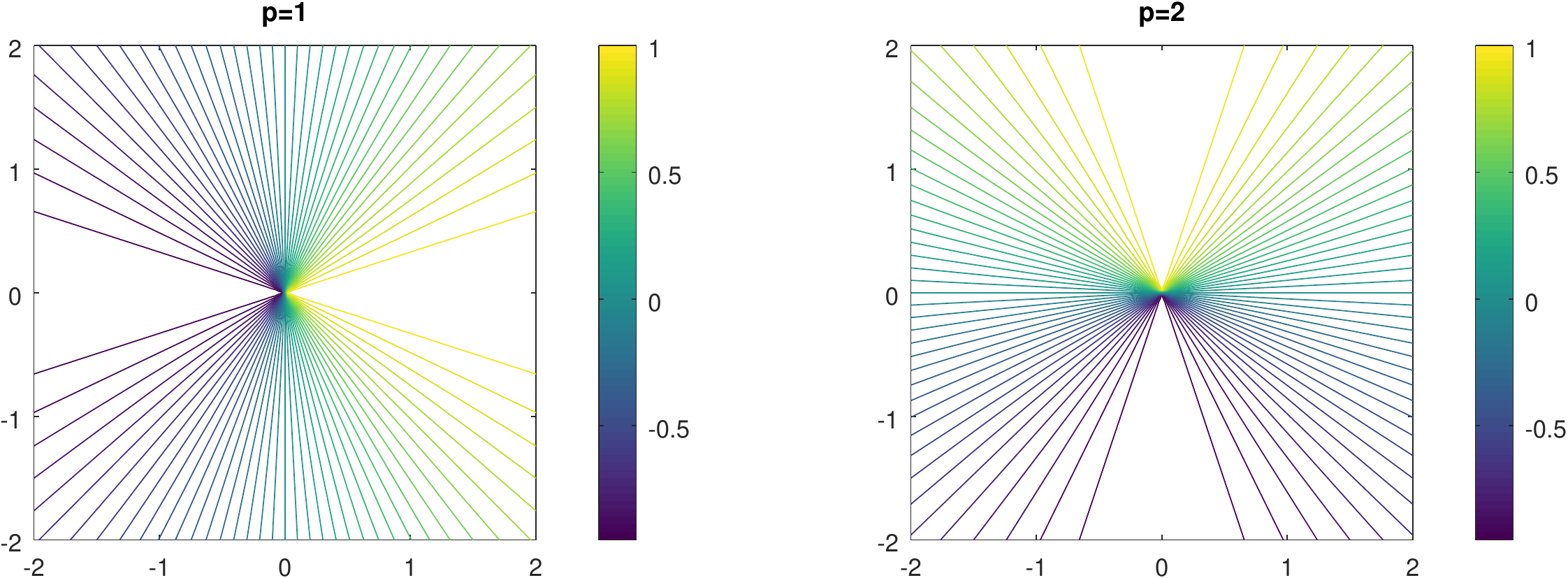}
\caption{The first order ABFs, derived from a conic RBF.\label{fig:abf1}}
\end{figure}

\begin{figure}
\includegraphics[width=\textwidth]{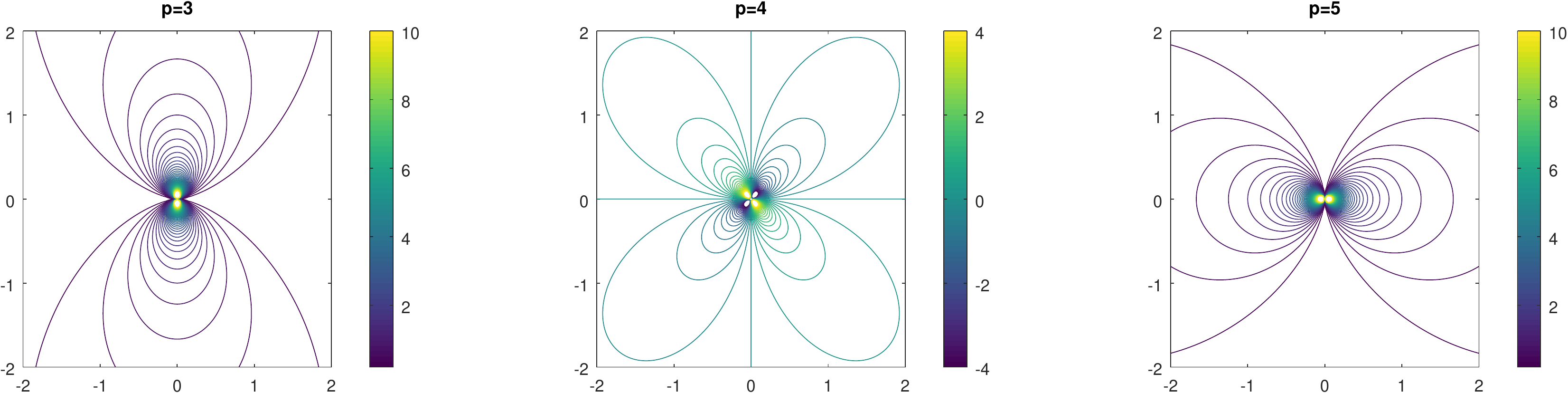}
\caption{The second order ABFs, derived from a conic RBF.\label{fig:abf2}}
\end{figure}

\begin{figure}
\includegraphics[width=\textwidth]{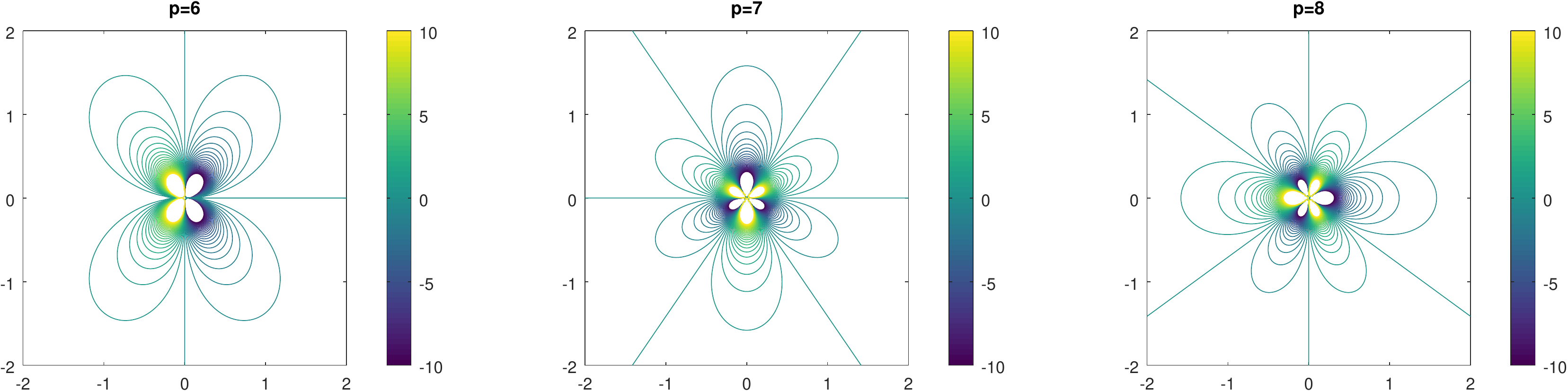}

~

\includegraphics[width=\textwidth]{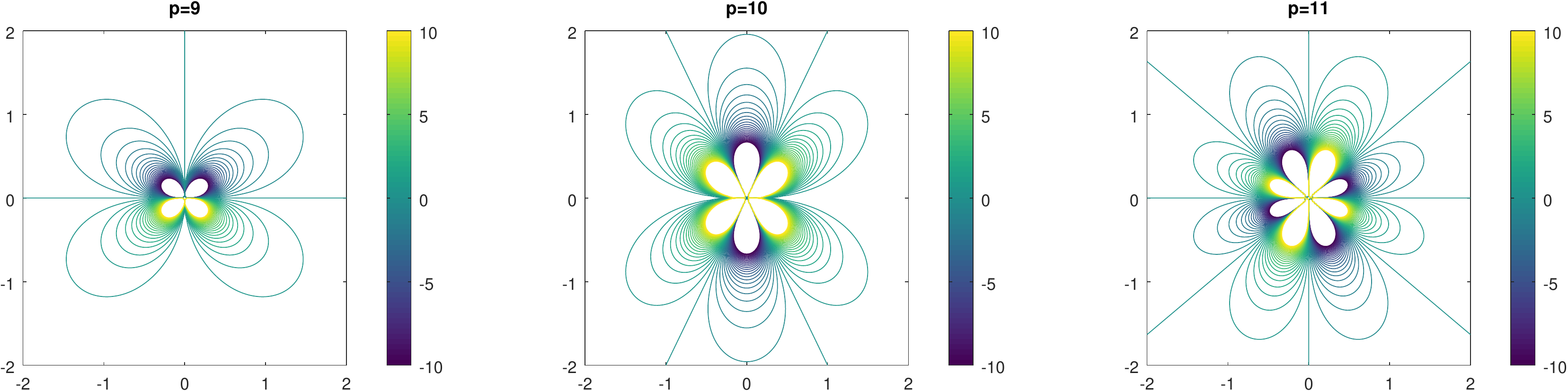}

~

\includegraphics[width=\textwidth]{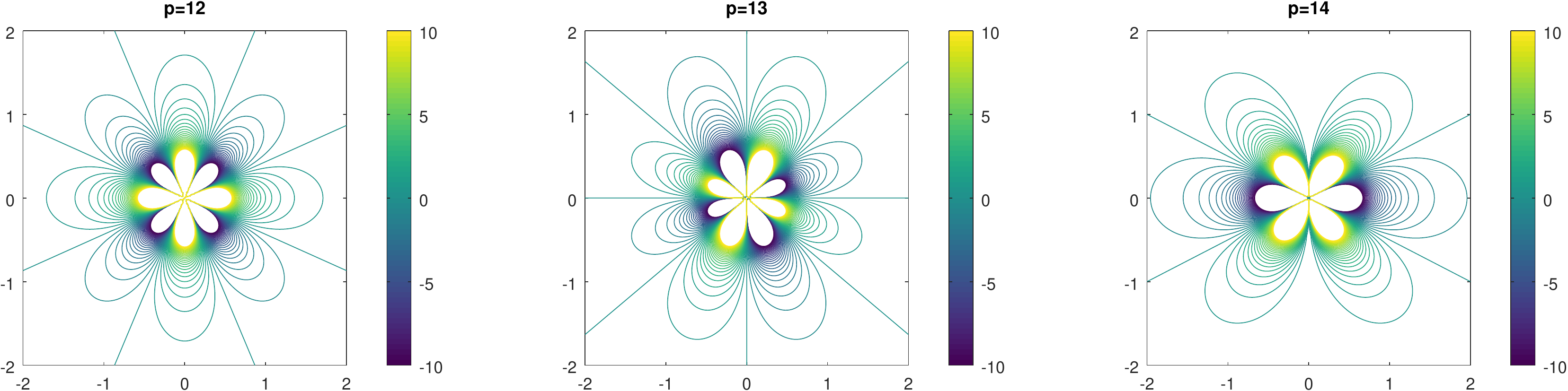}
\caption{The third and fourth order ABFs, derived from a conic RBF.\label{fig:abf34}}
\end{figure}

If the ABFs are chosen such that $\left\langle{W}^{m},X^{m}\right\rangle$ is independent of $h$ for all $m\le{p}$, where $p$ is the number of ABFs used, the approximation requirement~\eqref{eq:approx_req} will be satisfied. Given the elements of $\bm{X}_{ji}$ scale with $h$, $h^{2}$, $h^{3}$, $h^{4}$ and so on for $p\in\left[1,2\right]$, $p\in\left[3,5\right]$, $p\in\left[6,9\right]$, and $p\in\left[10,14\right]$ respectively, the ABFs must scale with $h^{-1}$, $h^{-2}$, $h^{-3}$, $h^{-4}$ and so on. This ensures that the diagonal of $\bm{M}$ is independent of resolution. This requirement may be satisfied by constructing the ABFs from the partial derivatives of some appropriately scaled fundamental RBF $W^{0}=W^{0}\left(\lvert\bm{r}_{ji}\rvert\right)$, according to
\begin{equation}\bm{W}_{ji}=\left.\bm{D}\left(W^{0}\right)\right\rvert_{\bm{r}_{ji}}\end{equation}
Taking $W^{0}$ as a very simple RBF - a cone - defined by
\begin{equation}W^{0}\left(q\right)=\frac{3}{4\pi}\left(1-\frac{q}{2}\right),\end{equation}
where $q=\left\lvert\bm{r_{ji}}\right\rvert/h=r_{ji}/h$, the radial derivative of $W^{0}$, $dW^{0}/dr_{j}=-3/8\pi{h}$, and higher radial derivatives are zero ($\partial^{k}W^{0}/\partial{r}_{j}^{k}=0$, $\forall{k}\ge{2}$). We obtain the vector of ABFs by applying the operator $\bm{D}$ to $W^{0}$:
\begin{multline}\bm{W}_{ji}=\left.\bm{D}\left(W^{0}\right)\right\rvert_{\bm{r}_{ji}}=\frac{-3}{8\pi{h}}\left[\frac{x}{r},\frac{y}{r},\frac{y^{2}}{r^{3}},\frac{-xy}{r^{3}},\frac{x^{2}}{r^{3}},\frac{-3xy^{2}}{r^{5}},\frac{2x^{2}y-y^{3}}{r^{5}},\frac{2xy^{2}-x^{3}}{r^{5}},\right.\\\left.\frac{-3x^{2}y}{r^{5}},\frac{12x^{2}y^{2}-3y^{4}}{r^{7}},\frac{9xy^{3}-6x^{3}y}{r^{7}},\frac{2x^{4}-11x^{2}y^{2}+2y^{4}}{r^{7}},\frac{9x^{3}y-6xy^{3}}{r^{7}},\frac{12x^{2}y^{2}-3x^{4}}{r^{7}}\dots\right]^{T},\label{eq:conic_abs}\end{multline}
in which the subscripts $ji$ have been dropped from $x_{ji}$, $y_{ji}$, and $r_{ji}$ for ease of exposition. Figures~\ref{fig:abf1},~\ref{fig:abf2}, and~\ref{fig:abf34} show the first, second, and third and fourth order ABFs respectively. We observe that because $\left.dW^{0}/dr_{j}\right\rvert_{q=0}\ne{0}$ (i.e. the fundamental RBF has a pointed peak), the ABFs generated from it have singularities at $\bm{r}_{ji}=0$. The first order ABFs (Figure~\ref{fig:abf1}) have a removable singularity. The second order ABFs have a simple pole, the third order ABFs have a second order pole, and the order $k$ ABFs have a pole of order $k-1$.

This approach may be used with an arbitrary fundamental RBF. In this work we investigate four choices of $W^{0}$: the cone, as described above, a quadratic RBF, the Wendland C6 kernel~\cite{wendland_1995}, and the Gaussian, details of which are provided in Appendix~\ref{rbf_details}. We note here that when we set $W^{0}$ to an SPH kernel with $p=5$, the matrix on the LHS of~\eqref{eq:lsys} is the transpose of the matrix obtained and solved in the method of~\citet{sibilla_2015}. 

\begin{figure}
\includegraphics[width=0.6\textwidth]{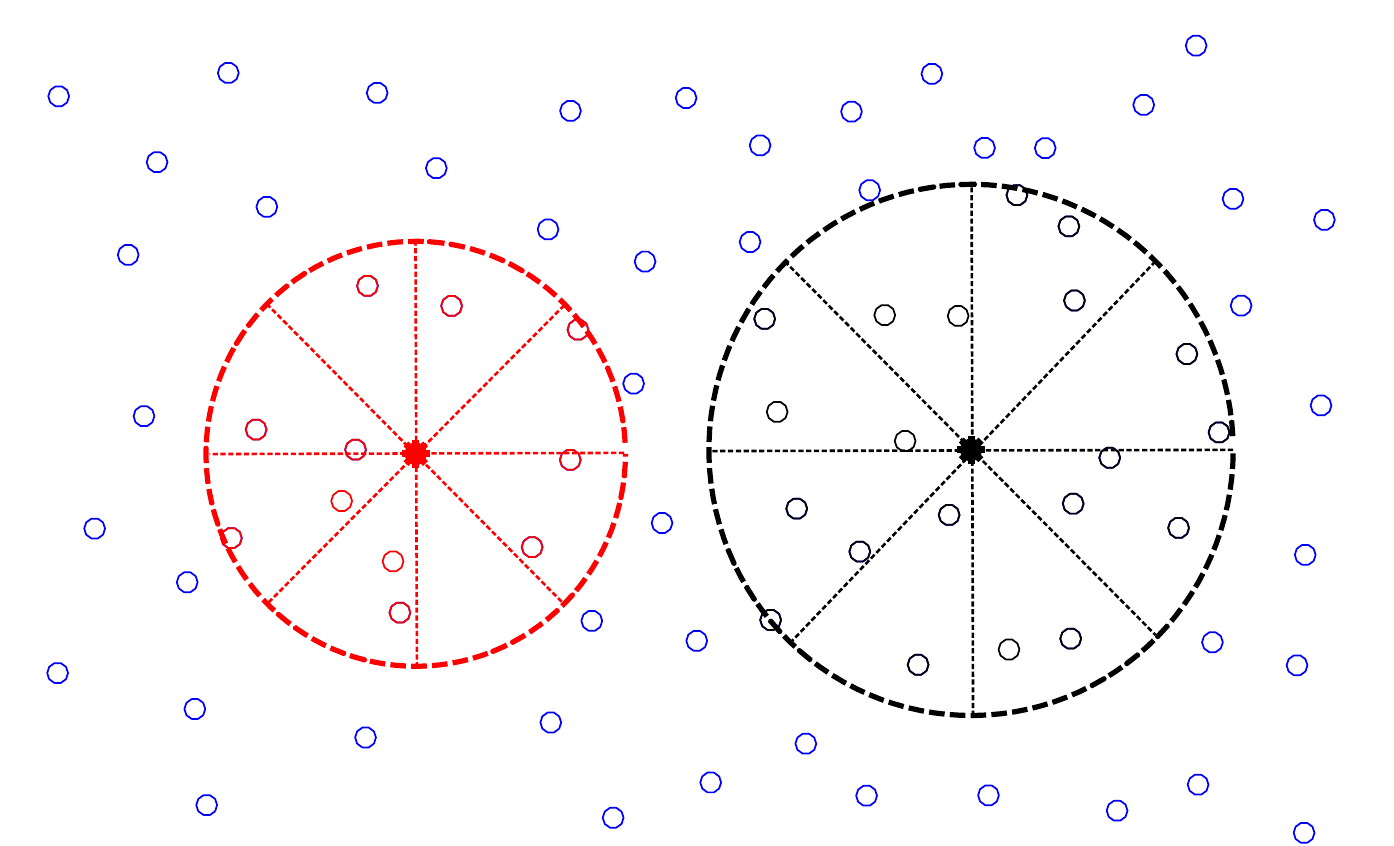}
\caption{Two example computational stencils. The red stencil (left) corresponds to $h/\delta{r}=1$, whilst the black stencil (right) corresponds to $h/\delta{r}=1.25$. The thin dashed lines divide the stencil into $2k$ equal segments for $k=4$.\label{fig:nyq1}}
\end{figure}

The statement in the previous section, that the node distribution must adequately sample the ABFs used is analagous to stating that the sums from which the elements of $\bm{M}_{i}$ are constructed must adequately approximate continuous integrals:
\begin{equation}\displaystyle\sum_{j}X^{m}_{ji}W^{n}_{ji}\approx\left\langle{X}^{m}_{ji},W^{n}_{ji}\right\rangle\qquad\forall{m,n}\in\left[1,p\right].\label{eq:adequately}\end{equation}
In practice this requirement imposes a lower limit on $h/\delta{r}$. For each value of $k$, there is a critical value of $h/\delta{r}$ (and hence $\mathcal{N}$) below which one or more of the eigenvalues of $\bm{M}_{i}$ has magnitude of machine precision zero, and below which the condition number of $\bm{M}_{i}$ increases significantly (typically by about 10 orders of magnitude). The critical values for $k=\left\{2,4,6,8\right\}$ are $h/\delta{r}_{crit}\approx\left\{0.75,1.15,1.6,2.1\right\}$ and in two dimensions $\mathcal{N}_{crit}\approx\left\{8,21,37,57\right\}$, and are (on average) independent of disorder of the node distribution. We note that the optimal number of nodes required for polynomial reconstruction of the same order is $\mathcal{N}_{poly}=\left\{6,15,28,45\right\}$ in two dimensions. It is clear from Figures~\ref{fig:abf1},~\ref{fig:abf2}, and~\ref{fig:abf34}, that as the order of ABF increases (higher $k$), more nodes are required, due to the increasing complexity of the ABFs (higher order poles, higher wavenumbers). When the ABFs in~\eqref{eq:conic_abs} are expressed in local polar coordinates ($r_{ji}$ and $\theta_{ji}$ having origin at $\bm{r}_{i}$), the order $k$ ABFs contain terms of $\cos{k\theta}$ and $\sin{k\theta}$. For the stencil to adequately sample the ABFs, the Nyquist sampling criteria must be satisfied. That is, if the domain of support is divided into $2k$ equal segments (each with angle $\pi/k$), there is a node in every segment. This is illustrated in Figure~\ref{fig:nyq1} for $k=4$. For $h/\delta{r}=1$ (left, red stencil) there is a segment containing no nodes: the $4^{th}$ order ABFs are undersampled by this stencil. With $h/\delta{r}=1.25$ (right, black stencil), all $8$ segments contain at least one node, and the stencil satisfies the Nyquist sampling criteria for the $4^{th}$ order ABFs. This criteria provides a more accurate definition of the word ``\emph{adequately}'' use for~\eqref{eq:adequately}, and the values of $h/\delta{r}_{crit}$ and $\mathcal{N}_{crit}$ correspond to the minimum values for which the Nyquist criteria is satisfied. The effects of choosing $h/\delta{r}<h/\delta{r}_{crit}$ can be seen in the numerical results in Section~\ref{conv}, in particular in the horizontal contours in Figure~\ref{fig:grad_sweep}.

\begin{figure}
\includegraphics[width=0.99\textwidth]{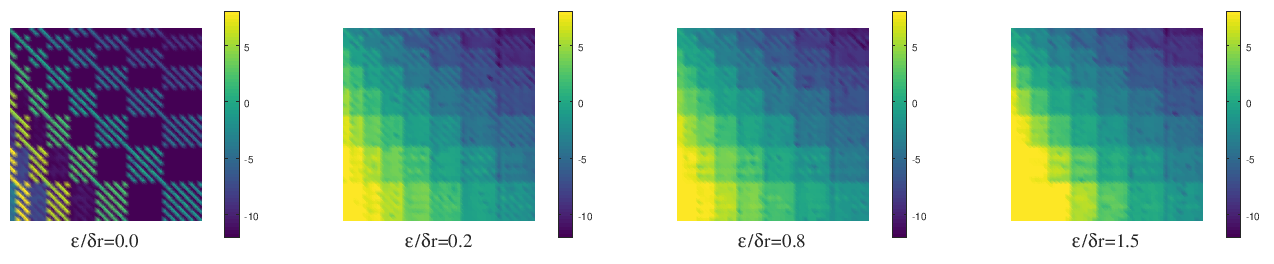}
\caption{Illustrations of the matrix $\bm{M}_{i}$ obtained with $k=8$ ($p=44$), and the quadratic ABFs, with $h/\delta{r}=2/5$, for increasingly noisy node distributions. Colour indicates the base $10$ logarithm of the absolute value of each element $\log_{10}\left(\lvert\bm{M}_{i}\rvert\right)$.\label{fig:matrix}}
\end{figure}

Figure~\ref{fig:matrix} illustrates an example of the matrix $\bm{M}_{i}$ for $k=8$ with the quadratic ABFs, $h/\delta{r}=2.5$, and a range of values of distribution noise $\varepsilon/\delta{r}$. We see for the uniform Cartesian distribution, $\bm{M}$ has a clear structure. The entries coloured in dark blue are zero \emph{to within machine precision}, due to the orthogonality of the odd order ABFs with the even order monomials (and vice versa). The magnitude of elements in top right are small, and bottom left are large, which is consistent with the scaling of $\bm{X}_{ji}$ and $\bm{W}_{ji}$ with $h$. For smaller $k$, the matrices $\bm{M}_{i}$ are just sub-matrices of those illustrated here. We see as the noise is increased, the magnitude of the entries in the bottom left corner increases (the brighter yellow patch), which is a consequence of the undersampling of the highest order ABFs. We investigate the effect of stencil size ($\mathcal{N}_{i}$ and $h/\delta{r}$) further in Section~\ref{sec:stab} below.

\section{Convergence and stability}\label{conv}

We now analyse the convergence properties of LABFM, and present a brief stability analysis.

\subsection{Convergence}
\begin{figure}
\includegraphics[width=0.49\textwidth]{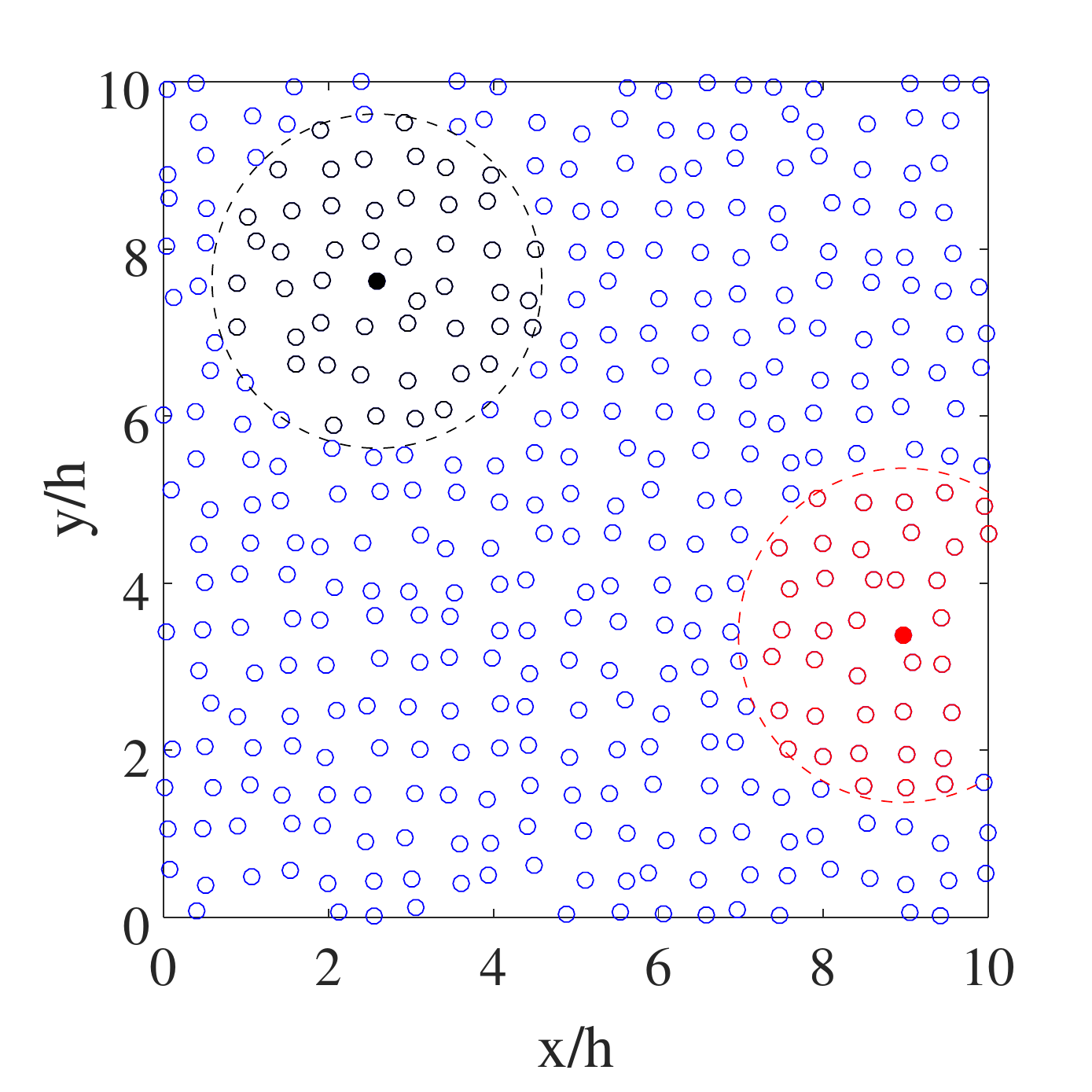}
\caption{An example node distribution with $\varepsilon/\delta{r}=0.5$, and $h/\delta{r}=2$. The dashed circles with radius $2h$ indicate the support domains of two nodes (indicated by filled circles). The computational stencil depicted in black has full support, and the computational stencil depicted in red has incomplete support.\label{fig:dist}}
\end{figure}

We consider a square domain defined by $\left(x,y\right)\in\left[0,H\right]\times\left[0,H\right]$. We define the test function
\begin{equation}\phi\left(\hat{x},\hat{y}\right)=1.0+\left(\hat{x}\hat{y}\right)^{4}+ \displaystyle\sum_{n=1}^{6}\left(\hat{x}^{n}+\hat{y}^{n}\right),\label{eq:p6}\end{equation}
where $\hat{x}=x-0.1453H$ and $\hat{y}=y-0.16401H$, and set $H=1$. This offset is psuedo-random, and included to ensure asymmetry in the function, to prevent the masking of errors which could cancel for a symmetric function. Nodes are distributed uniformly (Cartesian grid) with spacing $\delta{r}$, and randomly peturbed. The maximum peturbation distance is $\varepsilon$, and hence we take $\varepsilon/\delta{r}$ as a measure of the irregularity or noise of the distribution. In terms of fill distance and separation distance as used in the field of scattered data approximation, and defined in~\cite{wendland_2004}, $\varepsilon/\delta{r}=0$ corresponds to a separation distance of $\delta{r}/2$ and a fill distance of $\delta{r}/\sqrt{2}$. For $\varepsilon/\delta{r}=0.5$, the separation distance is approximately $0.29\delta{r}$, and the fill distance approximately $0.86\delta{r}$. Initially we set $h=2\delta{r}$ which, with a support radius of $2h$, yields $\mathcal{N}_{i}\approx{50}$ in two dimensions. An example node distribution and computational stencil is depicted in Figure~\ref{fig:dist}. The stencil indicated in black has full support, and the stencil indicated in red has incomplete support. At boundaries, we create additional ``ghost'' nodes (not shown in Figure~\ref{fig:dist}) with the same distribution properties in a strip of width $2h$ around the domain, ensuring that all computational nodes have fully supported stencils. We test our method by calculating approximations of the gradients and Laplacian of $\phi$ for a range of resolutions. As a measure of the error we take the normalised $L_{2}$-norm:
\begin{equation}L_{2}\text{-norm}\left(\cdot\right)^{d}=\frac{\left\{\displaystyle\sum_{i=1}^{N}\left[L^{d}_{i}\left(\cdot\right)-\bm{C^{d}}\cdot\left.\bm{D}\left(\cdot\right)\right\rvert_{i}\right]^{2}\right\}^{\frac{1}{2}}}{\left\{\displaystyle\sum_{i=1}^{N}\left[\bm{C^{d}}\cdot\left.\bm{D}\left(\cdot\right)\right\rvert_{i}\right]^{2}\right\}^{\frac{1}{2}}}\end{equation}

\begin{figure}
\includegraphics[width=0.49\textwidth]{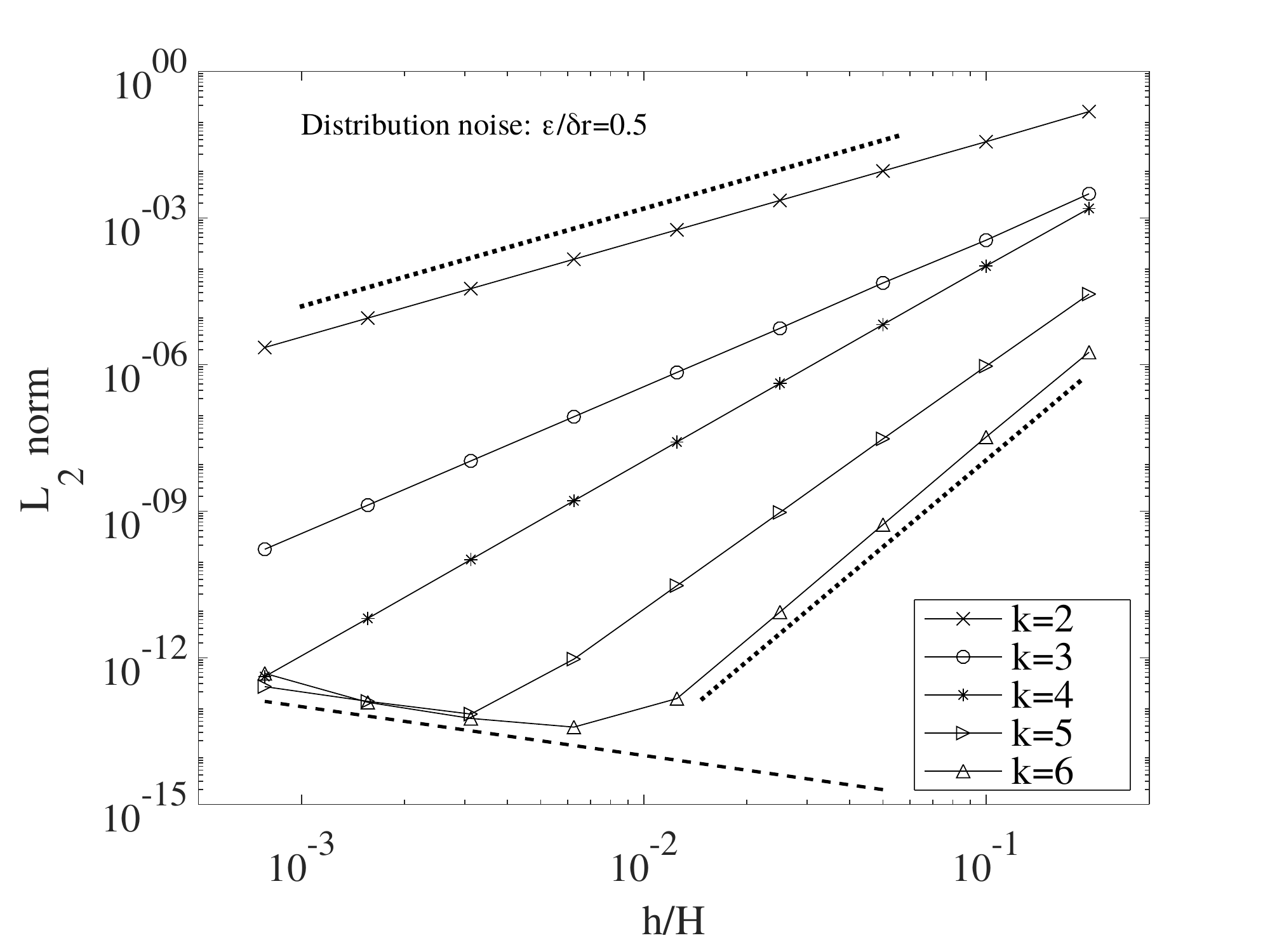}
\includegraphics[width=0.49\textwidth]{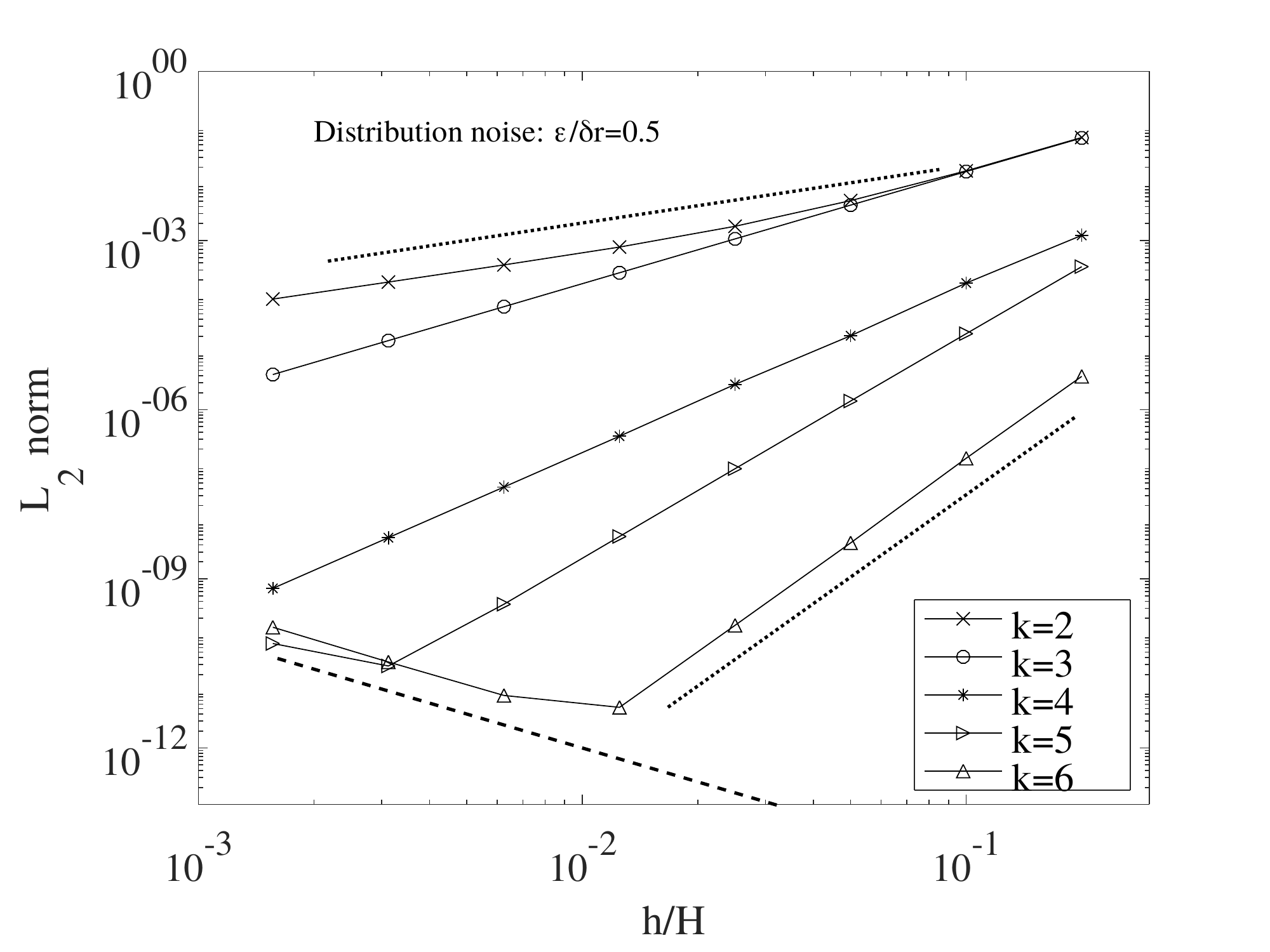}
\caption{Error in gradient (left) and Laplacian (right) approximations of~\eqref{eq:p6}. The dashed lines represent machine precision errors of $10^{-16}/h$ and $10^{-16}/h^{2}$ for gradients and Laplacians respectively. The dotted lines show convergence rates of $2$ and $6$ for gradients, and $1$ and $5$ for Laplacians.\label{fig:glerror}}
\end{figure}

Figure~\ref{fig:glerror} shows the $L_{2}$-norm of the error in gradient and Laplacian approximations for node distributions with $\varepsilon/\delta{r}=0.5$ (the distribution depicted in Figure~\ref{fig:dist}), for a range of resolutions, and increasing numbers of ABFs contributing to the discrete operator. For gradients, we observe orders of convergence increasing from second to sixth, as we increase $k$ from $2$ to $6$ ($p=5$ and $p=27$ respectively). The order of convergence for gradients is $k$. For the Laplacian, the order of convergence is $k-1$, ranging from first to fifth order. For $k=2$ there is a range of resolutions for which second order convergence is observed in the Laplacian approximation. However, this is because the second order error terms are larger than the first order error terms at these coarse resolutions, and this effect is strongly dependent on the choice of $\phi$. For $k>6$, we require $h/\delta{r}>2$. Setting $h/\delta{r}=2.5$, with $k=7$ we find convergence rates of $7$ and $6$ for gradients and Laplacian's respectively. With $k=8$ the errors in gradients are typically $10^{-14}$, and do not converge. This is because the $8^{th}$ order derivatives of $\phi$ (the highest non-zero derivatives) are constant. For Laplacian's with $k=8$ the errors scale with $h^{-1}$, but with magnitude $10^{-13}$ for the coarsest resolution. These errors are machine precision errors, accumulated during the construction of $\bm{M}$, and carried through the solution of~\eqref{eq:lsys}. For other test functions (e.g. a sinusoidally varying $\phi$) with $k=8$ the convergence rates of $8$ and $7$ for gradients and Laplacians hold. For clarity, we only show results were obtained using the conic ABFs in Figure~\ref{fig:glerror}. Quadratic ABFs would yield the same convergence rates, and error magnitudes approximately $32\%$ lower. 

\begin{figure}
\includegraphics[width=0.99\textwidth]{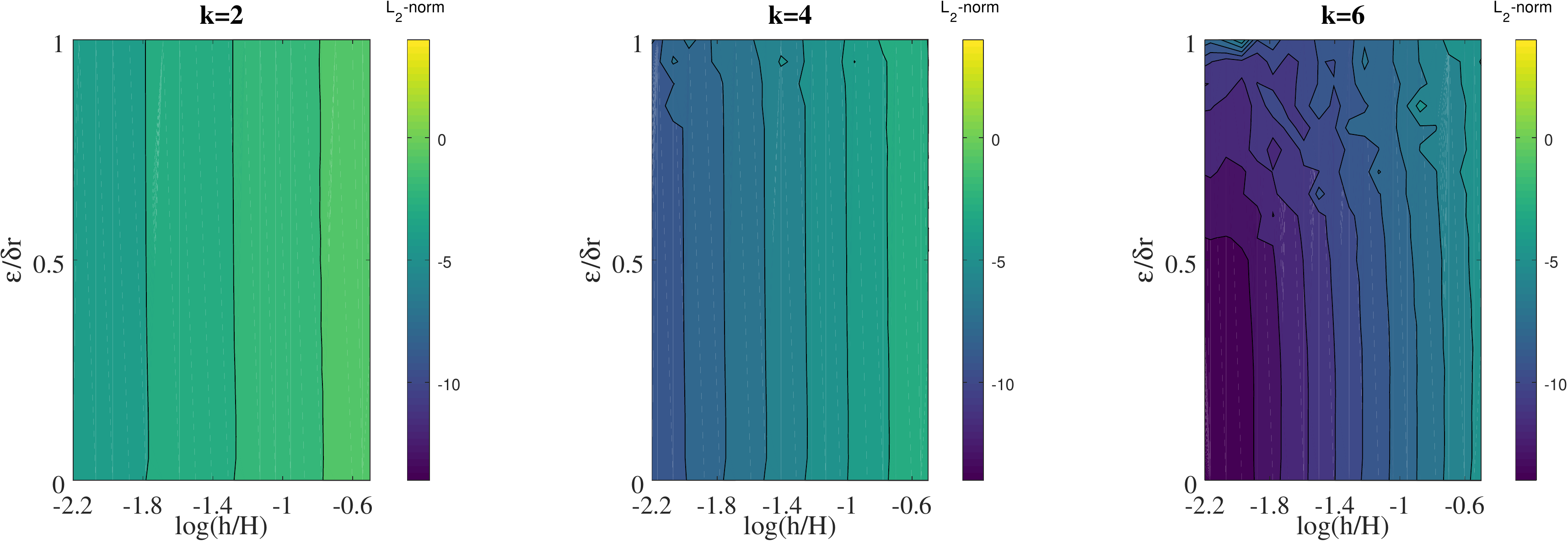}

~

\includegraphics[width=0.99\textwidth]{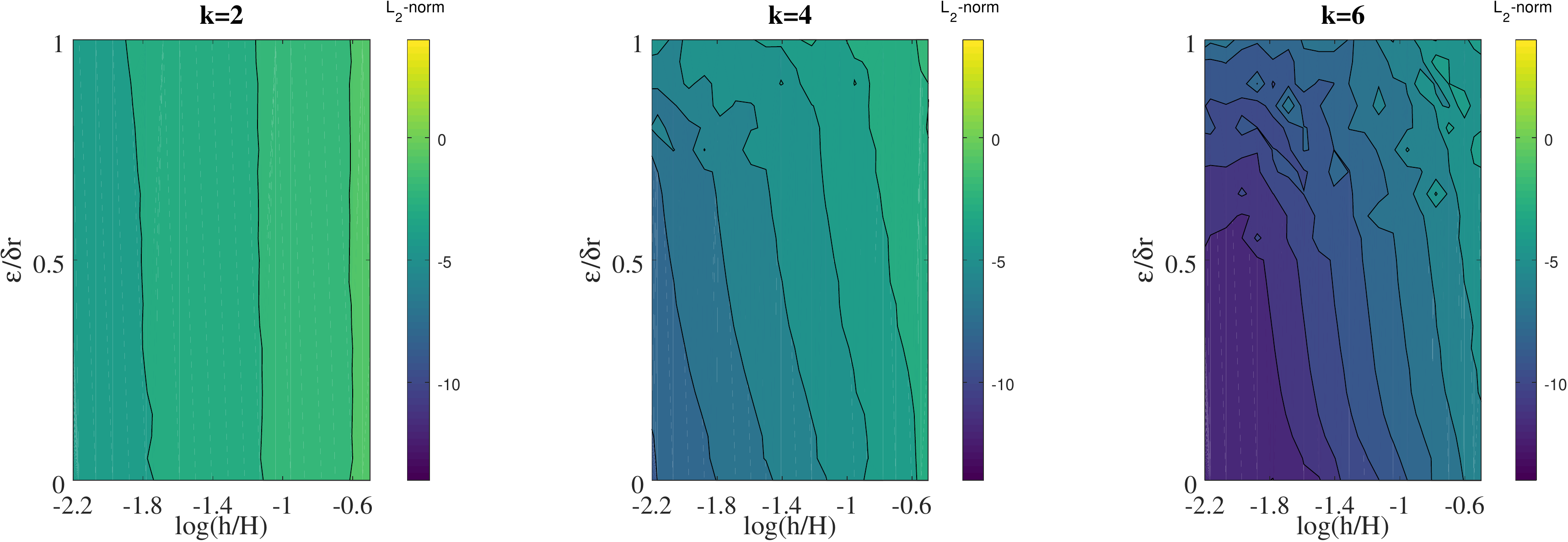}
\caption{$L_{2}$-norm when approximating the gradient (upper row) and Laplacian (lower row) of~\eqref{eq:p6}, for a range of resolutions and distribution noise, for $k=2$, $k=4$, and $k=6$. The scale on the abscissa is $\log_{10}\left(h/H\right)$, and colour indicates the base $10$ logarithm of the $L_{2}$-norm. The conic ABFs are used, with $h/\delta{r}=2$.\label{fig:noise_sweep}}
\end{figure}

Figure~\ref{fig:noise_sweep} shows the variation of the $L_{2}$-norm with resolution $h/H$ and the degree of noise in the node distribution $\varepsilon/\delta{r}$. For all $k$, we see a maintenence of convergence order (contours in Figure~\ref{fig:noise_sweep} are parallel) for levels of noise below $\varepsilon/\delta{r}=0.5$. For $k=2$ the convergence is maintained up to $\varepsilon/\delta{r}=1$ and beyond, whilst as $k$ increases, and convergence breaks down at lower $\varepsilon/\delta{r}$. The resilience of the method to noisy node distributions is greater for first derivatives (top row in Figure~\ref{fig:noise_sweep}) than for second derivatives (bottom row). For first derivatives, the magnitude of the error is also independent of $\varepsilon/\delta{r}$ as long as convergence is maintained (vertical contours), whilst for second derivatives, the magnitude of the error increases approximately linearly with increasing $\varepsilon/\delta{r}$ (contours are diagonal, and approximately straight). This result is intuitive, as all results in Figure~\ref{fig:noise_sweep} were obtained with the same $h/\delta{r}$, and so the sampling of the ABFs by the node distribution is \emph{relatively} better for $k=2$ than $k=6$.

\begin{figure}
\includegraphics[width=0.99\textwidth]{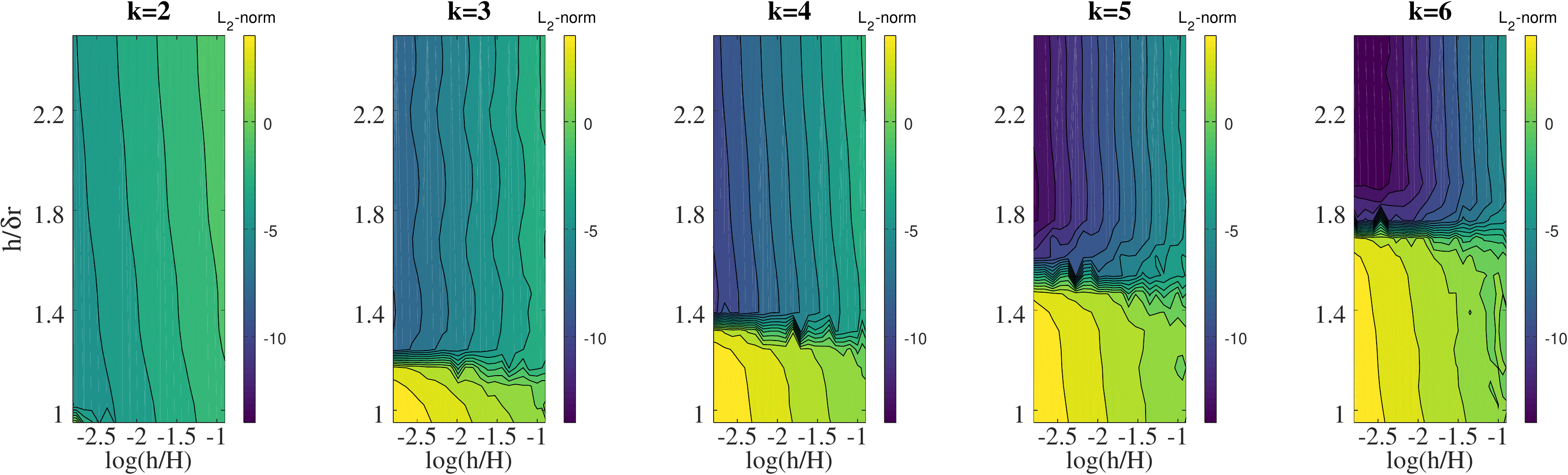}
\caption{Contour plots of the $L_{2}$-norm when approximating the gradient of~\eqref{eq:p6}, for a range of resolutions and stencil sizes ($h/\delta{r}\propto\sqrt{\mathcal{N}}$), for increasing values of $k$. The scale on the abscissa is $\log_{10}\left(h/H\right)$, and colour indicates the base $10$ logarithm of the $L_{2}$-norm. The node distribution is noisy Cartesian, with $\varepsilon/\delta{r}=0.5$.\label{fig:grad_sweep}}
\end{figure}

Figure~\ref{fig:grad_sweep} shows the variation of the $L_{2}$-norm with resolution $h/H$ and stencil size $h/\delta{r}\propto\sqrt{\mathcal{N}}$, as $k$ is increased from $2$ to $6$. We use $h/\delta{r}$ as a measure of stencil size because stencils are calculated based on a support radius ($2h$ here) as opposed to the desired stencil number size $\mathcal{N}$ as in RBF methods. For reference, a value of $h/\delta{r}=2$ corresponds to $\mathcal{N}\approx{50}$ in two dimensions. For all values of $k$, we see convergence of the $L_{2}$-norm with increasing resolution at large $h/\delta{r}$ (parallel contours in upper section of plots). However, there is a clear critical value of $h/\delta{r}$ below which the method is divergent. Below this value, the node distribution within the computational stencil does not adequately sample the ABFs, leading to ill conditioned matrices $\bm{M}_{i}$ in~\eqref{eq:lsys}. As discussed in the previous section, this critical value is larger for larger $k$, because for higher $k$ we include higher order ABFs which have a more complex structure, requiring a greater number of nodes to adequately sample them. In the majority of the remainder of this paper, we use $h/\delta{r}=2$, which leads to convergence for $k\le6$. However, we could, for low order schemes (e.g. $k=2$) use a significantly smaller stencils, with $h/\delta{r}=1$, reducing computational costs. For $k=8$, we observe $8^{th}$ order convergence with $h/\delta{r}=2.5$ ($\mathcal{N}\approx78$ in two dimensions). For a domain with incomplete support at the boundaries, we find the method yields accurate one sided derivatives and Laplacians, provided $h/\delta{r}$ (and hence, the computational stencil) is large enough at the domain boundaries. For nodes with incomplete support, the critical $h/\delta{r}$ is larger than in the fully supported case in Figure~\ref{fig:grad_sweep}. With this in mind, a potential approach when applying the method to practical problems is to have a region of higher resolution (smaller $\delta{r}$, maintiaining $h/\delta{r}$) in the vicinity of boundaries, to preserve the global convergence properties. 

\subsection{Stability analysis\label{sec:stab}}

\begin{figure}
\includegraphics[width=0.99\textwidth]{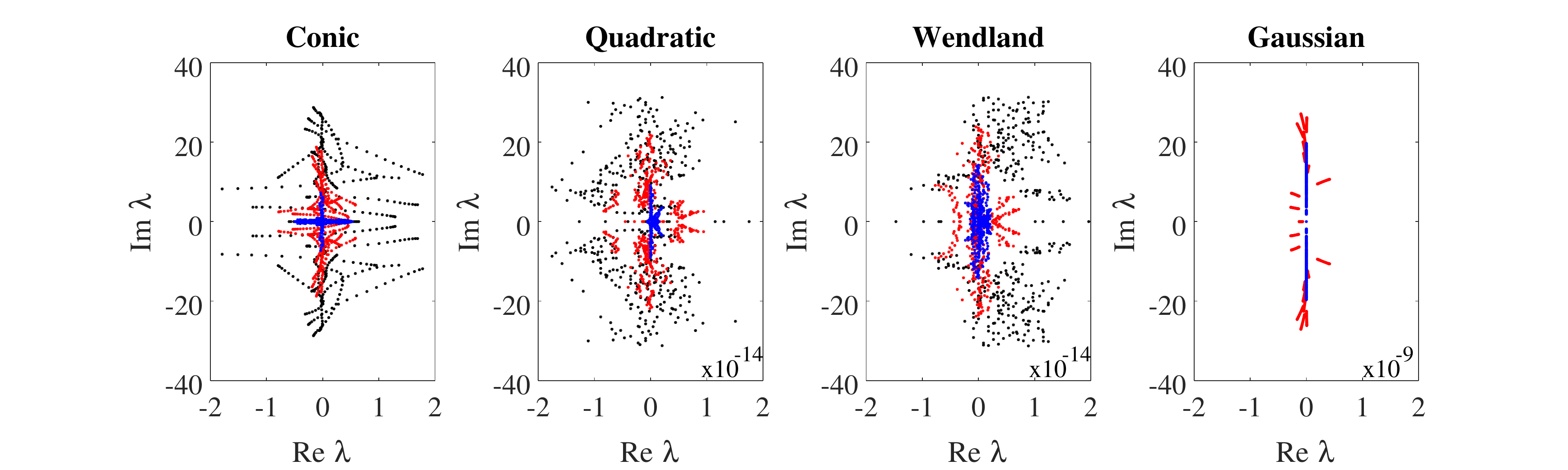}
\includegraphics[width=0.99\textwidth]{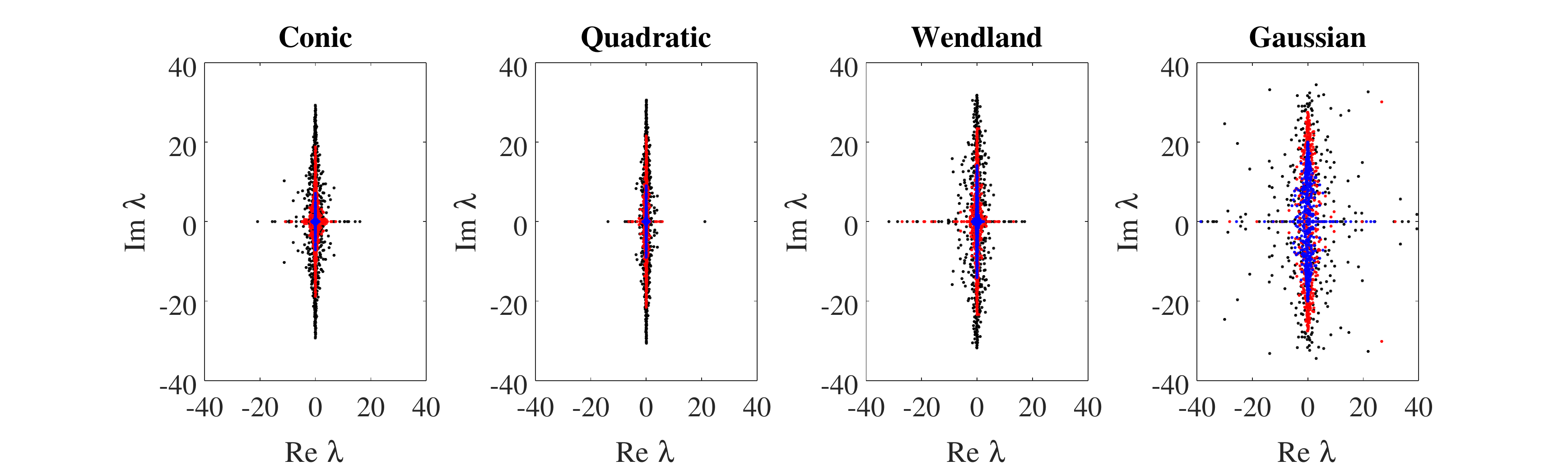}
\caption{The eigenvalues of the discretisation matrix $\bm{A^{x}}$ on a $441$ node Cartesian distribution, with boundaries providing complete support, for four sets of ABFs, generated from conic, quadratic, Wendland and Gaussian RBFs, and for $k=2$ (blue), $k=4$ (red), and $k=6$ (black). The upper row is for ideal Cartesian node distributions, whilst the lower row is for noisy Cartesian distributions, with $\varepsilon/\delta{r}=0.5$. Note the scalings of the abscissa in the upper row. In the right-most upper panel, the spread of the eigenvalues for $k=6$ is not shown, being approximately 3 orders of magnitude greater than that for $k=4$.\label{fig:eigens_convect}}
\end{figure}

To analyse the stability of the discrete operators, we construct a global derivative matrix $\bm{A^{d}}$ (following the procedure described in detail in Section~\ref{sec:poisson}). The eigenvalues of $\bm{A^{d}}$ provide information about the stability of the method. The convective derivatives (e.g. $d=x,y$ to approximate $\partial\left(\cdot\right)/\partial{x}$ or $\partial\left(\cdot\right)/\partial{y}$) are purely dispersive - that is they contain Fourier modes which correspond to a translation, with no growth or decay. Hence the eigenvalues $\lambda$ of the discrete convective operators $\bm{A^{x}}$ and $\bm{A^{y}}$ would (ideally) lie on imaginary axis ($\text{Re}\left(\lambda\right)=0$ $\forall\lambda$)~\cite{fornberg_2011}. For stability in purely convective problems, no modes should grow, and so $\text{Re}\left(\lambda\right)\le0$ $\forall\lambda$. Figure~\ref{fig:eigens_convect} shows the eigenvalues of $\bm{A^{x}}$ (with $N=121$ computational nodes, $h/\delta{r}=2$, and ghost nodes completing stencil support at boundaries) for Cartesian node distributions with and without noise, for ABFs generated from the Conic, Quadratic, Wendland, and Gaussian RBFs detailed in Appendix~\ref{rbf_details}. For ideal node distributions, the conic ABFs give $\bm{A^{x}}$ with substantial non-zero real parts to the eigenvalues, whilst for the quadratic, Wendland and Gaussian ABFs, the eigenvalues lie very close to the imaginary line, although the real parts of the eigenvalues using Gaussian ABFs are several orders of magnitude greater than the quadratic and Wendland ABFs (note the scalings in the upper row of Figure~\ref{fig:eigens_convect}). The scatter in the eigenvalues appears random (machine precision errors) with quadratic and Wendland ABFs, whilst there is some structure for the conic and Gaussian ABFs. Without proof, we postulate based on our numerical explorations that a condition for the real parts of the eigenvalues of $\bm{A^{x}}$ to be zero is that $dW^{0}/dr=0$ on the boundary of the support domain. This condition is exactly satisfied by the quadratic and Wendland ABFs, but not by the conic ABFs, and only approximately by the truncated Gaussian ABFs.

For noisy node distributions, the real parts of the eigenvalues of $\bm{A^{x}}$ become significant for all ABFs tested. We see that the conic and quadratic ABFs lead to eigenvalues lying closer to the imaginary axis than the Wendland and Gaussian ABFs, with the quadratic ABFs performing best, and Gaussian ABFs worst. For all distributions, and all choices of ABF, the eigenvalues of $\bm{A^{x}}$ fall closer to the imaginary axis for smaller $k$ (the blue dots in Figure~\ref{fig:eigens_convect} lie closer to $Im\left(\lambda\right)=0$ than the red and black dots.), suggesting that low order convective derivatives are likely to be more stable. 

\begin{figure}
\includegraphics[width=0.99\textwidth]{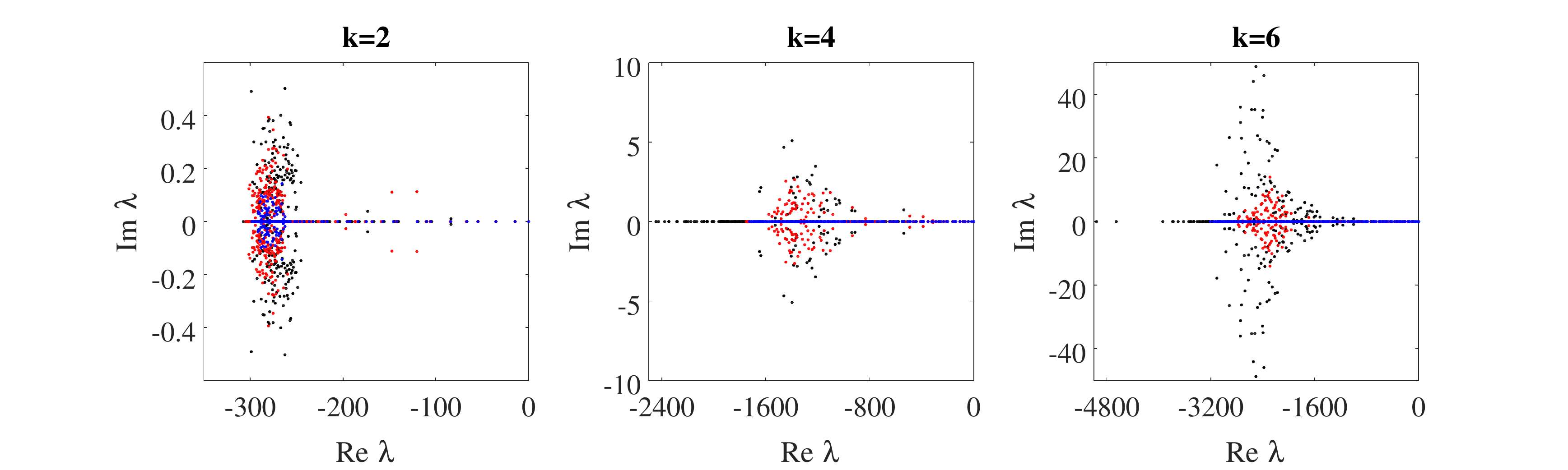}
\caption{The eigenvalues of the discretisation matrix $\bm{A^{L}}$ on a $441$ node noisy Cartesian distribution, with noise levels $\varepsilon/\delta{r}=0$ (blue), $\varepsilon/\delta{r}=0.2$ (red), and $\varepsilon/\delta{r}=0.5$ (black), for three values of $k$. In all cases, results are obtained with the Quadratic ABFs, and with $h/\delta{r}=2$.\label{fig:eigens_lap}}
\end{figure}

For the Laplacian, the eigenvalues of $\bm{A^{L}}$ should be as close as possible to the negative real axis, corresponding to decay of all modes. Figure~\ref{fig:eigens_lap} shows the eigenvalues of $\bm{A^{L}}$ for three levels of distribution noise, and three values of $k$, using the quadratic ABFs. For all $k$, the eigenvalues lie close to the real line, and the imaginary parts of the eigenvalues increase with increasing $\varepsilon/\delta{r}$. As $k$ is increased, the magnitude of the eigenvalues increases, and the spread of eigenvalues moves further along the negative real line. These results show that the Laplacian operator obtained with this method is stable, but that stability is conditional on the distribution noise. No modes grow in time (i.e. there are no eigenvalues with $Re\left(\lambda\right)>0$, and those modes which translate ($Im\left(\lambda\right)\ne0$) have significantly negative real parts to their eigenvalues, and so will be highly damped. The eigenvalue distribution when conic ABFs are used is similar to that shown in Figure~\ref{fig:eigens_lap}. For the Wendland and Gaussian ABFs, we find that $Re\left(\lambda\right)>0$ for at lower values of $\varepsilon/\delta{r}$. For each $k$ and $h/\delta{r}$, there is a certain noise level at which the Laplacian operator will become unstable ($Re\left(\lambda\right)>0$ for some $\lambda$). These values are listed in Table~\ref{tab:noise_limits}. We see that for small $k$, the Laplacian is stable for much greater noise levels than large $k$. For $k=2$, the Laplacian is stable for all $\varepsilon/\delta{r}$ tested. For large $k$, the Laplacian is stable for a range of $\varepsilon/\delta{r}$, provided a large enough $h/\delta{r}$ is chosen. Values of $\varepsilon/\delta{r}$ of $0.5$ or greater represent relatively severe distribution noise, and correspond to levels which can be easily obtained in complex geometries through iterative methods such as that described in Section~\ref{sec:poisson_annulus}. For Gaussian ABFs with $k=2$, there is an upper limit to $h/\delta{r}$ above which the Laplacian is unstable for all noise levels. In general the Gaussian ABFs have a smaller range of stable $h/\delta{r}$ and $\varepsilon/\delta{r}$ than the Wendland and quadratic ABFs, with the quadratic ABFs being stable for the greatest range of $\varepsilon/\delta{r}$. The conic ABFs (not shown in Table~\ref{tab:noise_limits}) exhibit similar maximum stable values of $\varepsilon/\delta{r}$ to the quadratic ABFs.

\begin{table}
\begin{center}
\caption{The maximum value of $\varepsilon/\delta{r}$ for stable $\bm{A^{L}}$ (i.e. $Re\left(\lambda\right)<0$ $\forall\lambda$) with the quadratic generated ABFs, on a $441$ node noisy Cartesian distribution. Entries of $>3$ indicate stability up the maximum value tested of $\varepsilon/\delta{r}=3$. Entries of ``none'' indicate there is no stable noise level.\label{tab:noise_limits}}
\begin{tabular}{l|c|c|c|c||c|c|c|c||c|c|c|c||}
~&\multicolumn{4}{c||}{\textbf{Quadratic}}&\multicolumn{4}{c||}{\textbf{Gaussian}}&\multicolumn{4}{c||}{\textbf{Wendland}} \\
$\bm{h/\delta{r}}$ & $\bm{k=2}$ & $\bm{k=4}$ & $\bm{k=6}$ & $\bm{k=8}$ & $\bm{k=2}$ & $\bm{k=4}$ & $\bm{k=6}$ & $\bm{k=8}$ &$\bm{k=2}$ & $\bm{k=4}$ & $\bm{k=6}$ & $\bm{k=8}$ \\
\hline
$\bm{1.2}$ & $1.8$ & $0.5$ & none & none & $0.5$ & $0.25$ & none & none & $0.85$ & $0.15$ & none & none\\
$\bm{1.6}$ & $>3$ & $0.9$ & $0.25$ & none & $0.55$ & $0.35$ & none & none & $1.1$ & $0.55$ & none & none \\
$\bm{2.0}$ & $>3$ & $1.0$ & $0.75$ & none & $0.2$ & $0.5$ & $0.1$ & none & $1.5$ & $0.7$ & $0.35$ & none\\
$\bm{2.4}$ & $>3$ & $1.1$ & $0.75$ & $0.4$ & none & $0.55$ & $0.2$ & none & $2.9$ & $0.75$ & $0.45$ & $0.1$\\
$\bm{2.8}$ & $>3$ & $1.1$ & $0.8$ & $0.4$ & none & $0.6$ & $0.35$ & none & $>3$ & $0.9$ & $0.7$ & $0.4$\\
\end{tabular}
\end{center}
\end{table}

For purely convective problems, all choices of ABF will be unstable on noisy node distributions, given the non-zero real parts to the eigenvalues of the convective derivative operators. A technique used to overcome this issue in RBF-FD is to introduce some hyperviscosity into the governing equations~\cite{fornberg_2011,flyer_2016_ns}, which provides the required dissipation to stabilise the solution. Hyperviscosity (i.e. $-\nabla^{4}$ or $\nabla^{6}$), is preferred to standard second order viscosity, as it acts more on shorter wavelength modes, which leaving long wavelength modes (relative to the node distribution) which are of physical interest, relatively untouched. The addition of hyperviscosity in the present method is straightforward. A discrete hyperviscosity operator $L^{hv}$ may be constructed in the same manner as the Laplacian and convective operators, by setting $\bm{C^{hv}}$ used in the RHS of~\eqref{eq:lsys} appropriately. The coefficient of hyperviscosity $\alpha$ scales with $h^{m}$, where $m$ is the order of the derivatives in the hyperviscosity operator, and is chosen large enough to ensure stability, but as small as possible. For example, to generate an $L^{hv}$ which approximates the biharmonic operator $-\alpha\nabla^{4}$ we set
\begin{equation}\bm{C^{hv}}=\begin{bmatrix}0,&0,&\dots&-\alpha,&0,&-2\alpha,&0,&-\alpha,&0&\dots\end{bmatrix}^{T},\end{equation}
where the non-zero elements are the elements $10$, $12$, and $14$. To include the triharmonic operator $\alpha\nabla^{6}$, we set elements $21$, $23$, $25$, and $27$ equal to $\alpha$, $3\alpha$, $3\alpha$, and $\alpha$ respectively. Note that in LABFM we can only create hyperviscosity operators up to $\nabla^{k}$. We test our hyperviscosity formulation in Section~\ref{burgers}.

\section{Application to Partial differential equations}\label{pde}
In this section we use LABFM to solve a range of PDEs. We have deliberately chosen prototype PDEs in unbounded or simple geometries in order to demonstrate the potential of the method for application to broad range of practical problems. Throughout this section we use ABFs generated from the quadratic RBF, unless otherwise specified. As in the previous section, we use the relative $L_{2}$-norm as a measure of error
\begin{equation}L_{2}\text{-norm}\left(\cdot\right)=\frac{\left\{\displaystyle\sum_{i=1}^{N}\left[\left(\cdot\right)^{LABFM}_{i}-\left(\cdot\right)^{exact}_{i}\right]^{2}\right\}^{\frac{1}{2}}}{\left\{\displaystyle\sum_{i=1}^{N}\left[\left(\cdot\right)^{exact}_{i}\right]^{2}\right\}^{\frac{1}{2}}},\end{equation}
where superscripts $LABFM$ and $exact$ indicate the numerical and analytic solutions respectively.

\subsection{Parabolic: Heat equation}

We first test the method on the prototypical parabolic equation - the (homegeneous) unsteady heat equation - given by
\begin{equation}\frac{\partial{u}}{\partial{t}}=\kappa\nabla^{2}u,\label{eq:heat}\end{equation}
where $\kappa$ is the coefficient of diffusivity. We solve~\eqref{eq:heat} on a periodic square domain $\left(x,y\right)\in\left[0,H\right]\times\left[0,H\right]$. With periodic boundary conditions, all computational nodes have complete stencils. The initial conditions (at time $t=0$) and solution are given by
\begin{equation}u\left(x,y,t\right)=\sin\left(\frac{2\pi{x}}{H}\right)\sin\left(\frac{2\pi{y}}{H}\right)\exp\left[\frac{-8\kappa\pi^{2}t}{H^{2}}\right].\end{equation}
We discretise the domain with a noisy Cartesian distribution, with the noise having magnitude $\varepsilon/\delta{r}=0.5$, where $\delta{r}$ is the noise-free node spacing. We set $h/\delta{r}=2$, and use the classical fourth order Runge-Kutta (RK4) scheme for time integration, and a time step of $\delta{t}=0.05h^{2}/\kappa$, setting $\kappa=1$.

\begin{figure}
\includegraphics[width=0.6\textwidth]{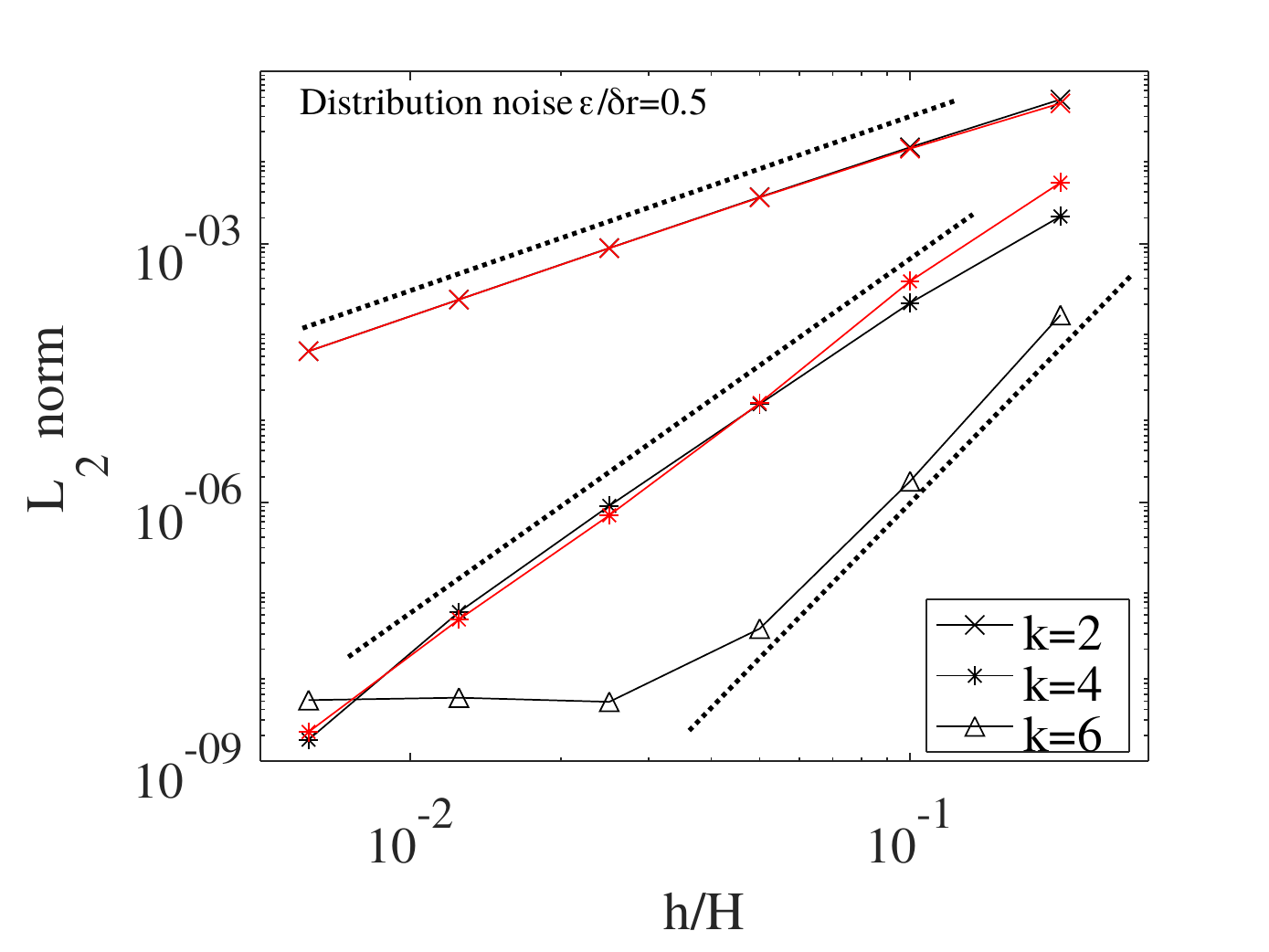}
\caption{Variation of the $L_{2}$-norm of the solution to the heat equation at $t^{\star}=1$ with resolution, for different values of $k$ (see legend). Black lines correspond to the case with periodic boundaries, and red lines indicate Dirichlet conditions with incomplete stencils near boundaries. The dotted lines correspond to convergence rates of $2$, $4$ and $6$.\label{fig:heat_conv}}
\end{figure}

The $L_{2}$-norm error of the numerical solution at non-dimensional time $t^{\star}=8\pi^{2}\kappa{t}/H^{2}=1$ is shown by the black lines and symbols in Figure~\ref{fig:heat_conv}. We see convergence rates of approximately $k$, as $k$ is varied between $2$ and $6$. Next, we replace the periodic boundary conditions with homogeneous Dirichlet conditions ($u=0$ on the boundary), so that the stencils of nodes near and on the boundaries are incomplete. Dirichlet conditions (either homogeneous or inhomogeneous) are imposed by prescribing $u$ for all nodes on the boundary, and solving~\eqref{eq:heat} only for internal nodes. The results with Dirichlet conditions are shown in red in Figure~\ref{fig:heat_conv}. For $k\le{4}$ the scheme is stable, and again we see convergence rates close to $k$. For $k>4$, the incomplete stencils near the boundaries (in particular the corners in the present case) lead to non-negative real parts to the eigenvalues of $\bm{A^{L}}$, and hence the scheme becomes unstable. We next replace the homogeneous Dirichlet boundary conditions with the inhomogeneous $u\left(x,0,t\right)=\sin\left(\pi{x}\right)$ for $x\in\left[0,H\right]$, and $u\left(x,y,t\right)=0$ on all other boundaries. Subject to these boundary conditions, the steady state solution of~\eqref{eq:heat} is given by
\begin{equation}u_{ss}\left(x,y\right)=\sinh\left(\pi\left[1-y\right]\right)\sin\left(\pi{x}\right)/\sinh\pi.\label{eq:heatss}\end{equation}
Table~\ref{tab:heatss} shows the $L_{2}$-norm and convergence rate of the steady state numerical solution (taken as the solution at non-dimensional time $t^{\star}=100$). We find the method converges with order between $k$ and $k+1$ (for $k\le4$). The ability to handle incomplete stencils accurately is a particularly attractive feature of LABFM. In effect, for $k\le{4}$ the method automatically generates one sided derivative approximations near boundaries, of order $k$. Indeed, for $k>4$, the gradient and Laplacian operators can still converge with order $k$, provided $\varepsilon/\delta{r}$ is small enough near boundaries, and $h/\delta{r}$ is large enough. Typically the range of acceptable $\varepsilon/\delta{r}$ and $h/\delta{r}$ decreases near boundaries with incomplete support. Obvious approaches to improving stability and accuracy near boundaries include increasing the resolution near boundaries, or decreasing $k$ to $4$ near boundaries - which will be explored in a future study on adaptivity.

\begin{table}
\begin{center}
\caption{The steady state $L_{2}$-norm error in the numerical solution of the heat equation subject to inhomogeneous boundary conditions (with analytical solution given by~\eqref{eq:heatss}). Figures in brackets show the convergence rate. Results were obtained using quadratic generated ABFs with $h/\delta{r}=2$ and $\varepsilon/\delta{r}=0.5$.\label{tab:heatss}}
\begin{tabular}{l|l|l|l|}
$\bm{h/H}$ & $\bm{k=2}$&$\bm{k=3}$& $\bm{k=4}$\\
\hline
$\bm{0.2}$ & $3.9\times{10}^{-3}$ (-) & $4.6\times{10}^{-4}$ (-) & $1.4\times{10}^{-4}$ (-) \\
$\bm{0.1}$ & $7.1\times{10}^{-4}$ ($2.46$)& $5.4\times{10}^{-5}$ ($3.08$) & $5.6\times{10}^{-6}$ ($4.68$) \\
$\bm{0.05}$ & $1.1\times{10}^{-4}$ ($2.73$) & $9.3\times{10}^{-6}$ ($2.54$) & $2.2\times{10}^{-7}$ ($4.67$)\\
$\bm{0.025}$ & $1.5\times{10}^{-5}$ ($2.83$) & $2.0\times{10}^{-6}$ ($2.20$) & $8.3\times{10}^{-9}$ ($4.72$) \\
$\bm{0.0125}$ & $2.2\times{10}^{-6}$ ($2.78$) & $4.7\times{10}^{-7}$ ($2.10$) & $2.9\times{10}^{-10}$ ($4.83$) \\
\end{tabular}
\end{center}
\end{table}

\subsection{Elliptic: Poisson's equation\label{sec:poisson}}

Our next PDE is Poisson's equation. Solving this elliptic PDE is in essence the inverse problem to the solution of the heat equation in the previous section. The Poisson's equation is given by
\begin{equation}\nabla^{2}\phi=f\left(x,y\right)\label{eq:poisson}\end{equation}
in the domain $\Omega$, with boundary conditions
\begin{subequations}
\begin{align}\phi=&g_{1}\left(x,y\right)\quad\text{on }\Gamma_{D}\label{eq:dir}\\\frac{\partial\phi}{\partial{n}}=&g_{2}\left(x,y\right)\quad\text{on }\Gamma_{N},\label{eq:neu}\end{align}\end{subequations}
where the boundary $\Gamma=\Gamma_{D}+\Gamma_{N}$. To solve~\eqref{eq:poisson} we construct a (linear) global discrete Laplacian operator $\bm{A^{d=L}}=\bm{A^{L}}$ (an $N\times{N}$ matrix) from the local operators $L_{i}^{L}$ (where the superscript $L$ indicates that the operator approximates the Laplacian). The $i$-th row of $\bm{A^{L}}$ is a rearrangment of the local operator $L_{i}^{L}$, which is be achieved by setting the elements of $\bm{A^{L}}$ as
\begin{subequations}
\begin{align}A^{L}_{i,j}&=w^{L}_{ji}\quad\forall{j}\ne{i}\\A^{L}_{i,i}&=-\displaystyle\sum_{j}w^{L}_{ji}.\end{align}
\end{subequations}
The discretised form of~\eqref{eq:poisson} is then
\begin{equation}\bm{A^{L}}\bm{\Phi}=\bm{F}\label{eq:poisson_system},\end{equation}
with solution vector 
\begin{equation}\bm{\Phi}=\begin{bmatrix}\phi_{1},&\phi_{2},&\dots,&\phi_{N}\end{bmatrix}^{T}\end{equation}
and source vector
\begin{equation}\bm{F}=\begin{bmatrix}f_{1},&f_{2},&\dots,&f_{N}\end{bmatrix}^{T}.\end{equation}
The linear system~\eqref{eq:poisson_system} may be solved using an iterative solver. In the present work, we use the stabilised bi-conjugate gradient algorithm, with a Jacobi preconditioner. Given the implicit method of solution, boundary conditions~\eqref{eq:dir} and~\eqref{eq:neu} must be incorporated into~\eqref{eq:poisson_system}, and this is done by manipulation of the rows, columns and elements of $\bm{A^{L}}$ and $\bm{F}$. On sections of the boundary $\Gamma_{D}$, we satisfy~\eqref{eq:dir} (on an example boundary node $b$) by setting
\begin{subequations}
\begin{align}A^{L}_{b,j}&=0\quad\forall{j}\ne{b}\\A^{L}_{b,b}&=1\\F_{b}&=g_{1,b},\end{align}\end{subequations}
which ensures that the solution to~\eqref{eq:poisson_system} yields $\phi_{b}=g_{1,b}$. On $\Gamma_{N}$, where Neumann boundary conditions are specified, we must satisfy both~\eqref{eq:poisson} and~\eqref{eq:neu} for each boundary node $b$. To do this, we use an approach which is consistent with the order of the spatial discretisation as follows. For every boundary node $b$ on $\Gamma_{N}$, we introduce an additional ``ghost'' node $\hat{b}$, such that $\bm{r}_{\hat{b}}=\bm{r}_{b}-\bm{n_{b}}\delta{r}$, where $\bm{n_{b}}=\left(n_{b,x},n_{b,y}\right)$ is the unit normal pointing into the domain at parent node $b$, and $\delta{r}$ is the initial node spacing. We increase the size of the linear system~\eqref{eq:poisson_system} to include the ghost nodes. The rows of $\bm{A^{L}}$ corresponding to ghost nodes $\hat{b}$ are used to enforce~\eqref{eq:neu}, by ensuring the discrete boundary normal derivative operator at node $b$ satisfies~\eqref{eq:neu}
\begin{equation}L^{n}_{b}=\displaystyle\sum_{j}\left(\cdot\right)_{jb}\left(n_{b,x}w^{x,bj}+n_{b,y}w^{y,bj}\right)=g_{2,b}.\end{equation}
We do this by setting, for every $\hat{b}$ and its parent $b$
\begin{subequations}
\begin{align}A^{L}_{\hat{b},j}&=n_{b,x}w^{x,bj}+n_{b,y}w^{y,bj}\quad\forall{j}\ne{b}\\A^{L}_{\hat{b},b}&=-\displaystyle\sum_{j}\left(n_{b,x}w^{x,bj}+n_{b,y}w^{y,bj}\right)\\F_{\hat{b}}&=g_{2,b},\end{align}
\end{subequations}
where the sum is over all $j$ which are neighbours of $b$. When the system~\eqref{eq:poisson_system} is solved with this modification, the elements of the solution vector corresponding to the ghost nodes ($\bm{\Phi}_{\hat{b}}$ $\forall\hat{b}$) will take values which ensure that the Neumann boundary condition~\eqref{eq:neu} is satisfied on $\Gamma_{N}$. We note that this approach to Neumann boundary conditions appears (coincidentally) similar to that described by~\citet{mishra_2019}.

\subsubsection{Periodic test case}

\begin{figure}
\includegraphics[width=0.99\textwidth]{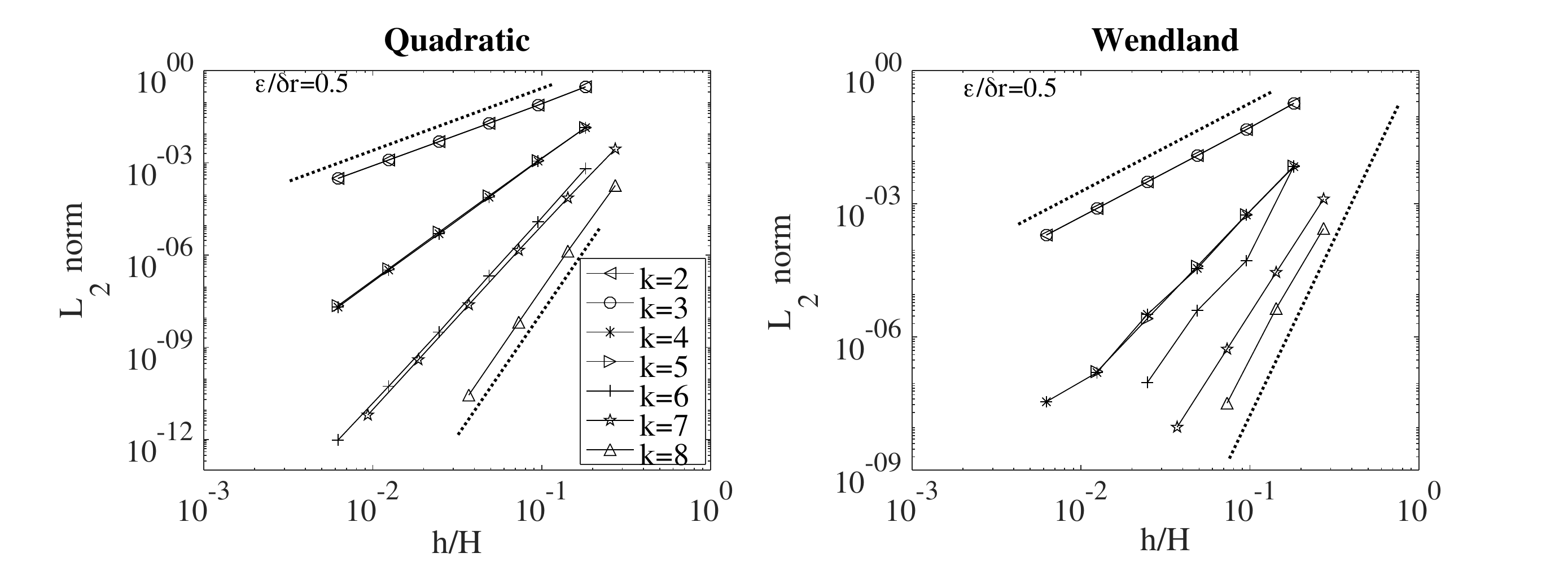}
\caption{Variation of the $L_{2}$-norm with resolution for the periodic Poisson's equation case 1 (periodic), for $k\in\left[2,8\right]$. The left panel shows errors using ABFs obtained from the Quadratic RBF, and the right panel shows ABFs obtained from the Wendland RBF. In all cases, $\varepsilon/\delta{r}=2$. For $k\le6$ we use $h/\delta{r}=2$, and for $k=7,8$ we use $h/\delta{r}=3$. The dotted lines show convergence rates of $2$ and $8$.\label{fig:poisson_periodic_conv}}
\end{figure}

Our first test case for Poisson's equation uses a periodic square domain $\left(x,y\right)\in\left[0,H\right]\times\left[0,H\right]$, and sets $f$ in~\eqref{eq:poisson} such that $\phi=\sin\left(2\pi{x}/H\right)\sin\left(2\pi{y}/H\right)$. Figure~\ref{fig:poisson_periodic_conv} shows the $L_{2}$-norm of the solution for a range of resolutions and values of $k$. Using the Quadratic ABFs (left panel) we find convergence orders of $2$ for $k=2,3$, $4$ for $k=4,5$, $6$ for $k=6,7$ and $8$ for $k=8$. For the $k=8$ the linear solver fails to converge at the finest resolutions. With the Wendland ABFs (right panel) the convergence rates follow the same pattern as with the Quadratic ABFs, although but at higher $k$, the solution becomes dependent on the node distribution disorder. The linear solver fails to converge at the finest resolution for $k\ge5$. This improved stability of the Quadratic over the Wendland ABFs is consistent with the observation in~\ref{sec:stab} that the eigenvalues of $\bm{A^{L}}$ obtained with Wendland ABFs contain positive real parts for lower levels of distribution noise than when Quadratic ABFs are used.

\subsubsection{Annular domain test case\label{sec:poisson_annulus}}
\begin{figure}
\includegraphics[width=0.6\textwidth]{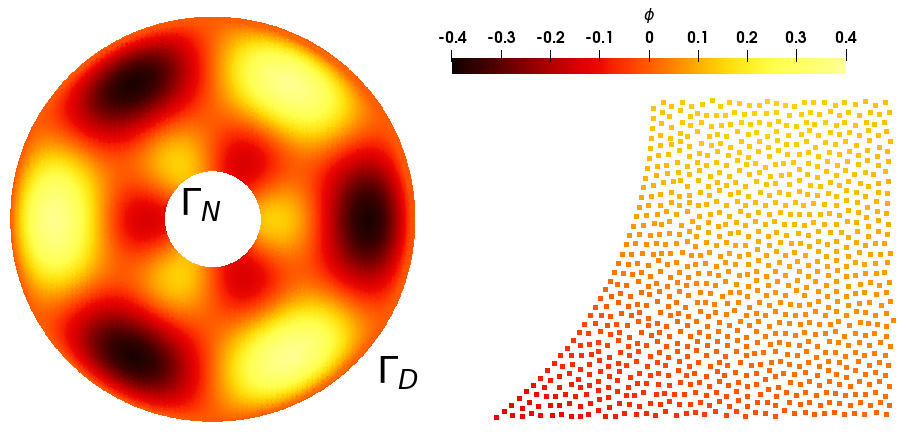}
\caption{The domain and solution for Poisson problem 2 (left), with a close up of a typical node distribution (right).\label{fig:poisson_sol}}
\end{figure}

Next we use an annular domain with inner diameter $H/4$ and outer diameter $H$, centred on the origin. The domain is illustrated in the left panel of Figure~\ref{fig:poisson_sol}. We solve~\eqref{eq:poisson} with the source term
\begin{equation}f=\left[12\pi\cos{4\pi{r}}-\left(16\pi^{2}r-\frac{1}{r}\right)\sin{4\pi{r}}\right]\cos{3\theta}-\frac{9}{r}\cos{3}\theta\sin{4\pi{r}},\end{equation}
with $r$ nd $\theta$ being polar coordinates based at the origin, subject to the boundary conditions $g_{1}=0$ on $\Gamma_{D}$ (the external boundary) and $g_{2}=\cos{3}\theta$ on $\Gamma_{N}$ (the internal boundary). The solution is given by
\begin{equation}\phi=r\sin{4\pi{r}}\cos{3\theta},\end{equation}
and shown in the left panel of Figure~\ref{fig:poisson_sol}. The node distribution is generated by discretising the boundaries (placing nodes at intervals $\delta{r}$ along all boundaries, then filling the domain with a noisy Cartesian distribution, with $\varepsilon/\delta{r}=0.5$. Temporary ghost nodes are generated outside the domain to complete support, whilst an iterative procedure is used to shift nodes according to:
\begin{equation}\bm{r}^{n+1}_{i}=\bm{r}^{n}_{i}+\frac{\delta{r}^{2}}{h}\displaystyle\sum_{j\in\left\lvert\bm{r}_{ji}\right\rvert<h}\left(\frac{\left\lvert\bm{r}_{ji}\right\rvert}{h}-1\right)\frac{\bm{r}_{ji}}{\left\lvert\bm{r}_{ji}\right\rvert}.\label{eq:iter_shift}\end{equation}
We apply typically $10$ iterations of~\eqref{eq:iter_shift} to obtain the final node distribution, an example of which is shown in Figure~\ref{fig:poisson_sol} (right panel). The procedure given by~\eqref{eq:iter_shift} moves nodes from regions of high node density to low node density, providing a more uniform distribution. The temporary ghost nodes prevent nodes from being shifted out of the domain during the application of~\eqref{eq:iter_shift}. After the shifting procedure, the temporary ghost nodes are discarded, and a ghost node is added for every node on $\Gamma_{N}$ as described above. The procedure~\eqref{eq:iter_shift} is relatively unrefined, however the resilience of LABFM to node distribution noise allows us to use this cheap procedure with only a few iterations, and still obtain high order convergence rates.

Table~\ref{tab:poisson_annular} shows the variation of the $L_{2}$-norm of the solution with resolution, for $k=\left\{2,3,4\right\}$. For $k=2,3$, we observe second order convergence, whilst for $k=4$ we observe fourth order convergence. As with the parabolic problem in the previous section, for $k>4$, the ABFs are inadequately sampled on nodes near the boundaries, where computational stencils have incomplete support. For $k\le4$, LABFM automatically generates one sided differences as required. A similar result was found by~\citet{bayona_2017}, who observed that when solving elliptic problems with RBF-FD, the method could handle incomplete support provided large enough computational stencils were used near boundaries. For $k>4$, the linear solver used to solve~\eqref{eq:poisson} did not converge. Preliminary investigations in which we set $k=4$ near boundaries, and $k>4$ on internal nodes were promising, yielding stable solutions, but for the present problem, only a minor increase in accuracy. This \emph{adaptive accuracy} approach will be explored further in a future study.

\begin{table}
\begin{center}
\caption{The $L_{2}$-norm error in the numerical solution of Poisson's equation case 2 (annular), for different $k$. Figures in brackets show the convergence rate. Results were obtained using quadratic generated ABFs with $h/\delta{r}=2$.\label{tab:poisson_annular}}
\begin{tabular}{l|l|l|l|}
$\bm{h/H}$ & $\bm{k=2}$&$\bm{k=3}$& $\bm{k=4}$\\
\hline
$\bm{2/13}$ & $0.655$ (-) & $0.539$ (-) & $1.05$ (-) \\
$\bm{2/25}$ & $0.136$ ($2.40$)& $0.146$ ($2.00$) & $0.101$ ($3.58$) \\
$\bm{2/49}$ & $3.5\times{10}^{-2}$ ($2.01$) & $3.7\times{10}^{-2}$ ($2.03$) & $2.9\times{10}^{-3}$ ($5.28$)\\
$\bm{2/97}$ & $9.1\times{10}^{-3}$ ($1.98$) & $9.5\times{10}^{-3}$ ($2.01$) & $1.4\times{10}^{-4}$ ($4.44$) \\
$\bm{2/193}$ & $2.3\times{10}^{-3}$ ($1.98$) & $2.4\times{10}^{-3}$ ($2.00$) & $6.5\times{10}^{-6}$ ($4.47$) \\
$\bm{2/385}$ & $5.9\times{10}^{-4}$ ($1.99$) & $6.0\times{10}^{-4}$ ($2.00$) & $3.6\times{10}^{-7}$ ($4.20$) \\
\end{tabular}
\end{center}
\end{table}


\subsection{Hyperbolic-Parabolic: Viscous Burgers' equation\label{burgers}}

Next we test LABFM on the viscous Burgers' equation, which is of mixed hyperbolic-parabolic type, and a prototypical shock-admitting conservation equation. The viscous Burgers' equation is written
\begin{equation}\frac{\partial\bm{u}}{\partial{t}}+\bm{u}\cdot\nabla\bm{u}=\frac{1}{Re}\nabla^{2}\bm{u}\label{eq:burgers}
\end{equation}
where $Re$ is the Reynolds number, and $\bm{u}=\left(u,v\right)$ is a velocity vector. We integrate~\eqref{eq:burgers} using the RK4 scheme, with a time step of 
\begin{equation}\delta{t}=\min\left(0.2h/\max\lvert{u}\rvert,0.05h^{2}Re\right)\label{eq:tstep}.\end{equation} 
We solve~\eqref{eq:burgers} for two cases.

\subsubsection{Travelling wave test case\label{non_steepening}}
\begin{figure}
\includegraphics[width=0.9\textwidth]{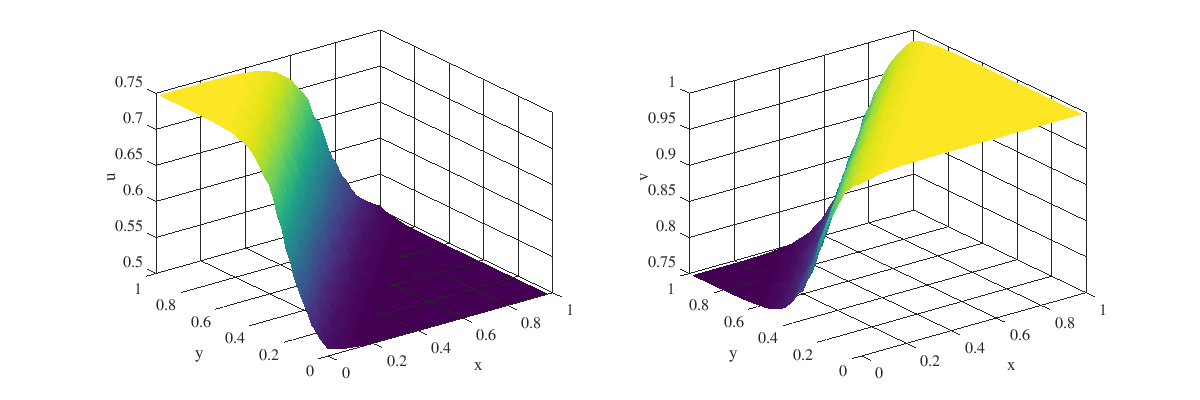}
\caption{Numerical solution to the Burgers' equation case 1 (travelling wave) at non-dimensional time $t=1$ for $Re=100$. Left and right panels show $u$ and $v$ respectively.\label{fig:burgers_tw_sol}}
\end{figure}

\begin{figure}
\includegraphics[width=0.6\textwidth]{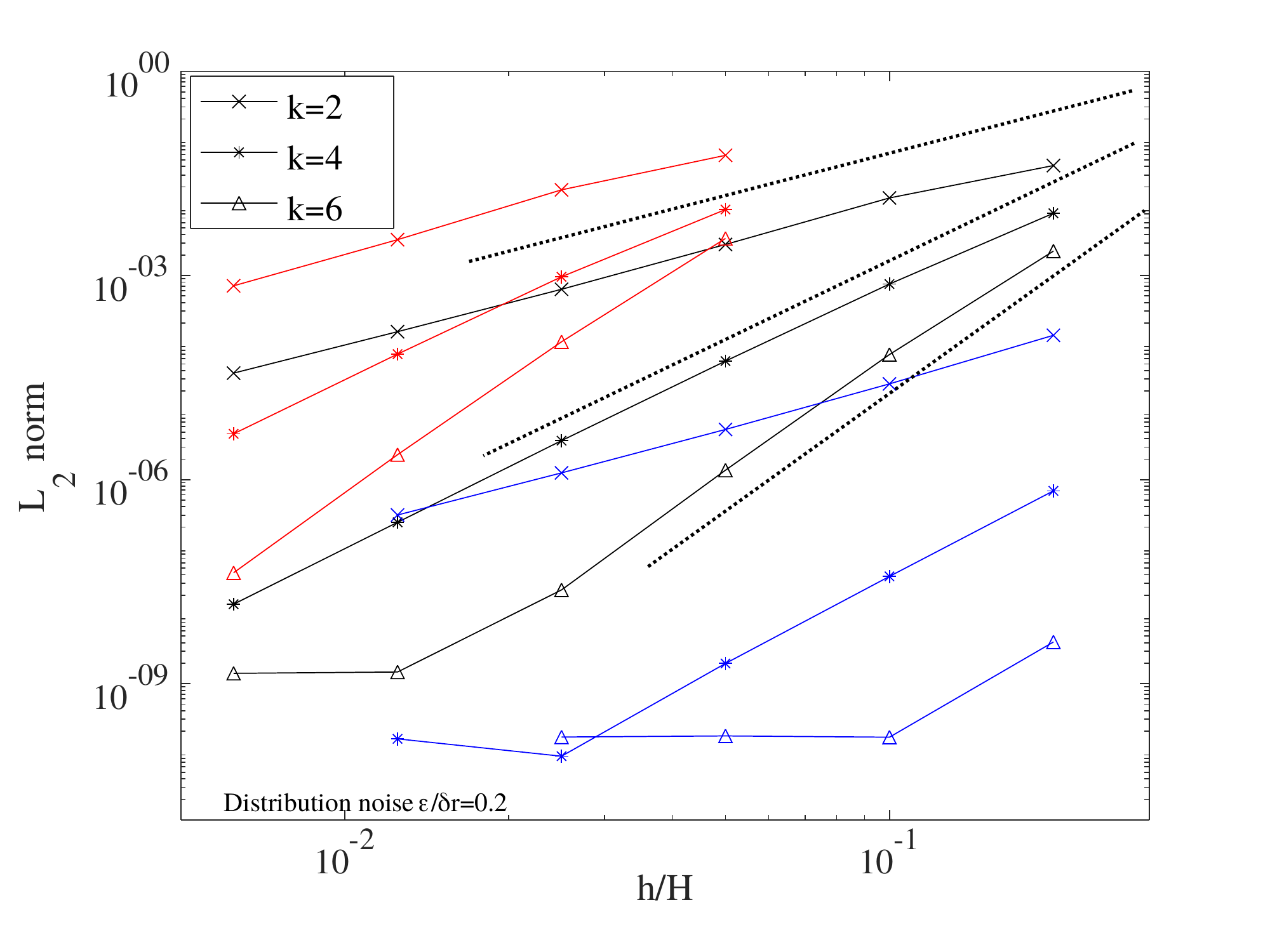}
\caption{$L_{2}$-norm of the error in the solution of Burgers' equation case 1 (travelling wave) at non-dimensional time $t=1$, for a range of resolutions, for various $k$ (see symbols in legend), and different Reynolds numbers: $Re=10$ (blue lines), $Re=100$ (black lines) and $Re=500$ (red lines). In all cases, $h/\delta{r}=2$ and $\varepsilon/\delta{r}=0.2$. The dotted lines show convergence rates of $2$, $4$ and $6$.\label{fig:burgers1_conv}}
\end{figure}

The first test comprises a travelling wave oriented diagonally along the $xy-$plane, in which steepening due to non-linear advection is balanced by viscous diffusion. The steepness of the travelling wave increases with increasing Reynolds number, and in the inviscid limit the wave becomes a discontinuity. We use a computational domain $\left(x,y\right)\in\left[0,1\right]\times\left[0,1\right]$, discretised with a noisy Cartesian node distribution with $\varepsilon/\delta{r}=0.2$. At the boundaries, we continue the node distribution outwards to provide full support, and the analytic solution is prescribed on nodes outside the boundary. The initial conditions and solution are
\begin{subequations}\begin{align}u\left(x,y,t\right)&=\frac{3}{4}-\frac{1}{4\left(1+e^{Re\left(-t-4x+4y\right)/32}\right)},\\v\left(x,y,t\right)&=\frac{3}{4}+\frac{1}{4\left(1+e^{Re\left(-t-4x+4y\right)/32}\right)}.\end{align}\end{subequations}
The numerical solution at $t=1$ is shown in Figure~\ref{fig:burgers_tw_sol}. Figure~\ref{fig:burgers1_conv} shows the convergence of errors for this problem at $t=1$, for increasing $k$, for three Reynolds numbers, $Re=10$ (blue lines), $Re=100$ (black lines), and $Re=500$ (red lines). For the highest Reynolds number the shock is too steep to be resolved by the coarsest two resolutions (not shown), and the simulations are unstable. The magnitude of the errors increase with increasing $Re$, due to the increasingly steep gradients present in the solution. For all cases, the solution converges with order approximately $k$, and convergence is within $\left[k-1/2,k+1/2\right]$. For $Re=10$ and $k=6$, the coarsest resolution shown corresponds to only $10\times10$ over the domain, and yet the relative error is less than $10^{-8}$.
 
\subsubsection{Periodic test case\label{sawtooth}}

\begin{figure}
\includegraphics[width=0.6\textwidth]{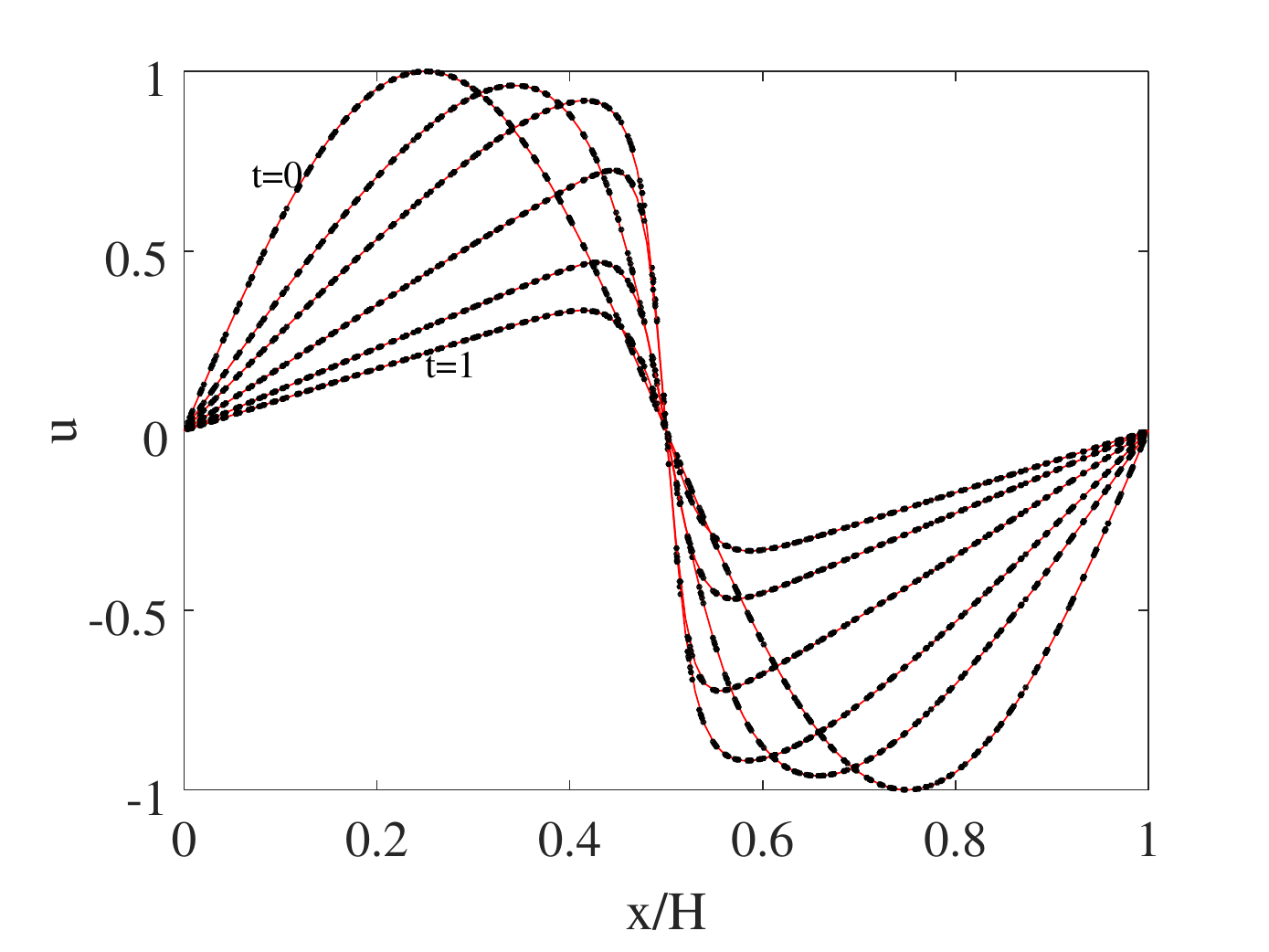}
\caption{The analytic (red line) and numerical solutions (black dots) of the periodic Burgers' problem with $Re=100$, $h/\delta{r}=2$, and $\varepsilon/\delta{r}=0.5$, at times $t=0$, $t=0.1$, $t=0.2$, $t=0.4$, $t=0.7$ and $t=1$, obtained with a resolution of $h/H=2/81$.\label{fig:burgers_sol1}}
\end{figure}

\begin{figure}
\includegraphics[width=0.99\textwidth]{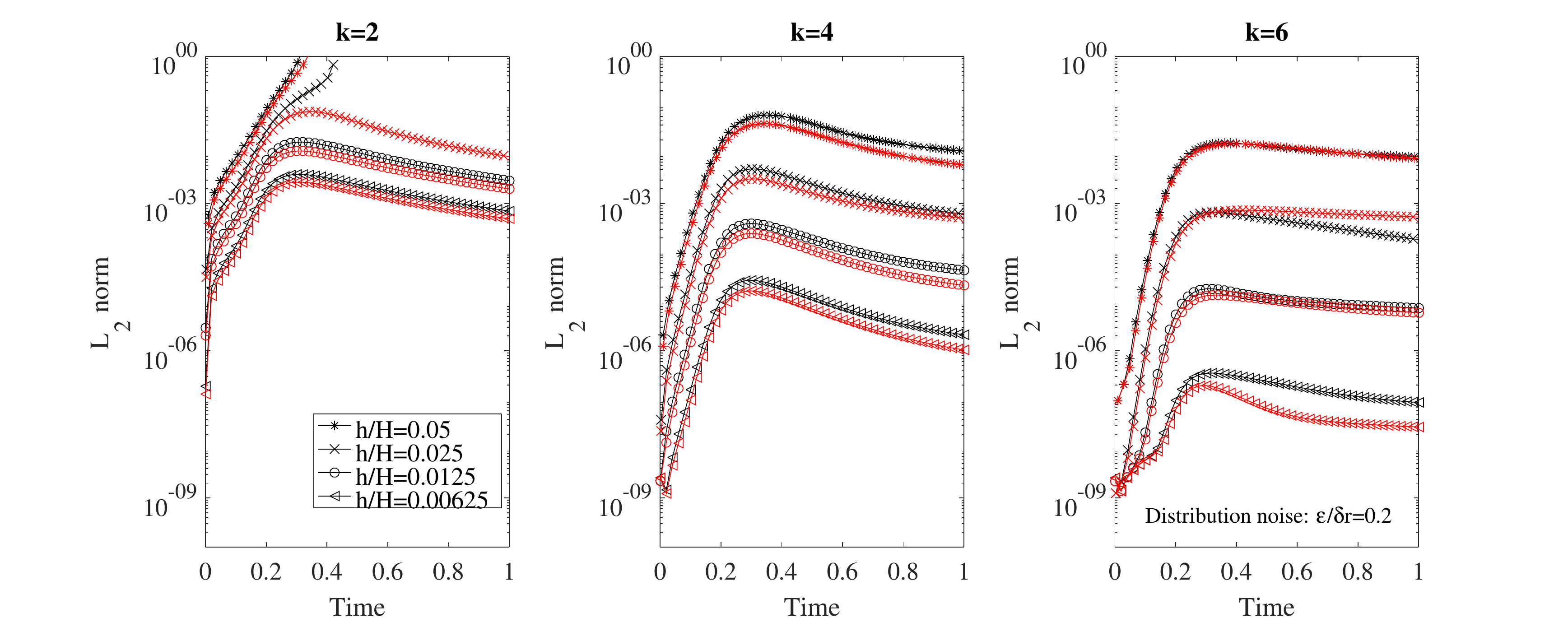}
\caption{Variation of the $L_{2}$-norm of the solution to the periodic Burgers' problem with time, for a range of resolutions (identified in legend), for the conic ABFs (black) and quadratic ABFs (red). The node distribution is noisy Cartesian, with $\varepsilon/\delta{r}=0.2$.\label{fig:burgers_l2time}}
\end{figure}

Next we solve~\eqref{eq:burgers} on a periodic square domain $\left(x,y\right)\in\left[0,H\right]\times\left[0,H\right]$, with sinusoidal initial conditions: $u\left(x,y,0\right)=\sin\left(2\pi{x}/H\right)$ and $v\left(x,y,t\right)=0$. The advection terms contribute to the development of a shock (as the sinusoidal solution tends to a sawtooth wave), which the viscous terms dissipate. The solution was first derived by~\citet{cole_1951}, and has since been used as a benchmark for numerical methods by numerous authors, for example~\cite{sibilla_2015}. The solution is given by
\begin{equation}u\left(x,t\right)=\frac{4\pi}{H{Re}}\frac{\displaystyle\sum_{n=1}^{\infty}nA_{n}\sin\left(\frac{2n\pi{x}}{H}\right)e^{\left(\frac{-n^{2}\pi^{2}t}{H^{2}Re}\right)}}{A_{0}+\displaystyle\sum_{n=1}^{\infty}A_{n}\cos\left(\frac{2n\pi{x}}{H}\right)e^{\left(\frac{-n^{2}\pi^{2}t}{H^{2}Re}\right)}}\end{equation}
where 
\begin{equation}A_{0}=\exp\left[\frac{-ReH}{4\pi}\right]I_{0}\left(\frac{ReH}{4\pi}\right);\quad{A}_{n}=2\exp\left[\frac{-ReH}{4\pi}\right]I_{n}\left(\frac{ReH}{4\pi}\right),\end{equation}
and $I_{n}$ are modified Bessel functions of the first kind. Note that $v\left(x,y,t\right)=0$ for all $t$. We use $Re=100$, and find that only the first $30$ terms are necessary, as $A_{n}<10^{-16}$ for $n\geq30$. Figure~\ref{fig:burgers_sol1} shows the analytic (red lines) and numerical solutions (black dots) at a range of times, from $t=0$ to $t=1$, with $\varepsilon/\delta{r}=0.5$, $k=4$, $h/\delta{r}=2$ and $h/H=2/81$. Figure~\ref{fig:burgers_l2time} shows the variation of the $L_{2}$-norm with time, for a range of resolutions and $k$, with $\varepsilon/\delta{r}=0.2$. As with the previous case, we found convergence rates between $k$ and $k-1$ as expected. These convergence rates held when the disorder in the node distribution was increased to $\varepsilon/\delta{r}=0.5$. The quadratic ABFs (red symbols) provide slightly better accuracy than the conic ABFs (black). For all simulations which remain stable, the errors grow during the advection dominated initial stages (as the sinusoidal profile steepens), and then decrease during the viscosity dominated stages thereafter. For $k=2$, the coarsest resolution is unstable for both conic and quadratic ABFs, and the second coarest resolution is unstable for the conic ABFs (left panel of Figure~\ref{fig:burgers_l2time}). Generally we observed the quadratic ABFs to be slightly more accurate, and slightly more stable (for this problem) than the conic ABFs.

\begin{figure}
\includegraphics[width=0.49\textwidth]{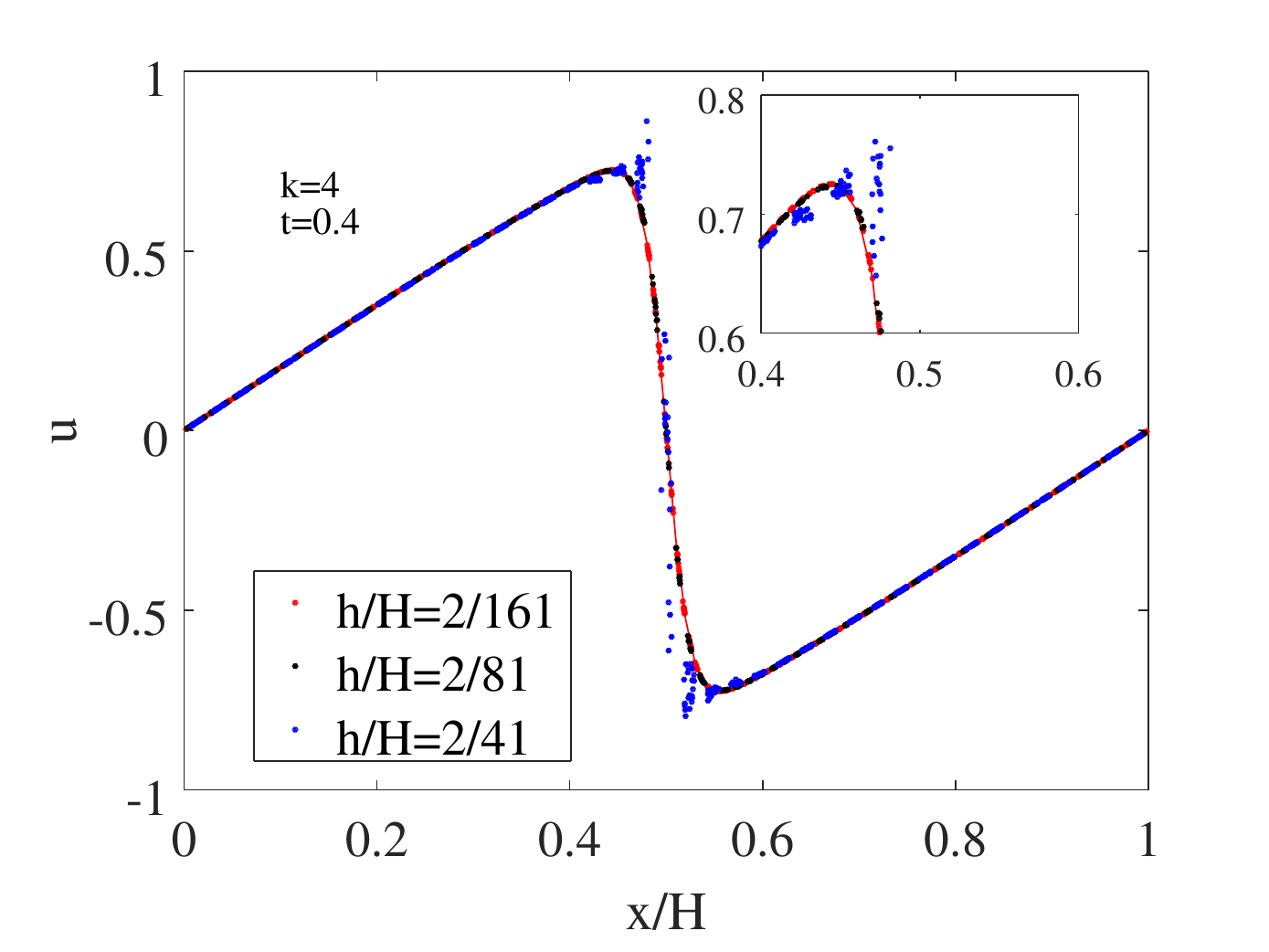}
\includegraphics[width=0.49\textwidth]{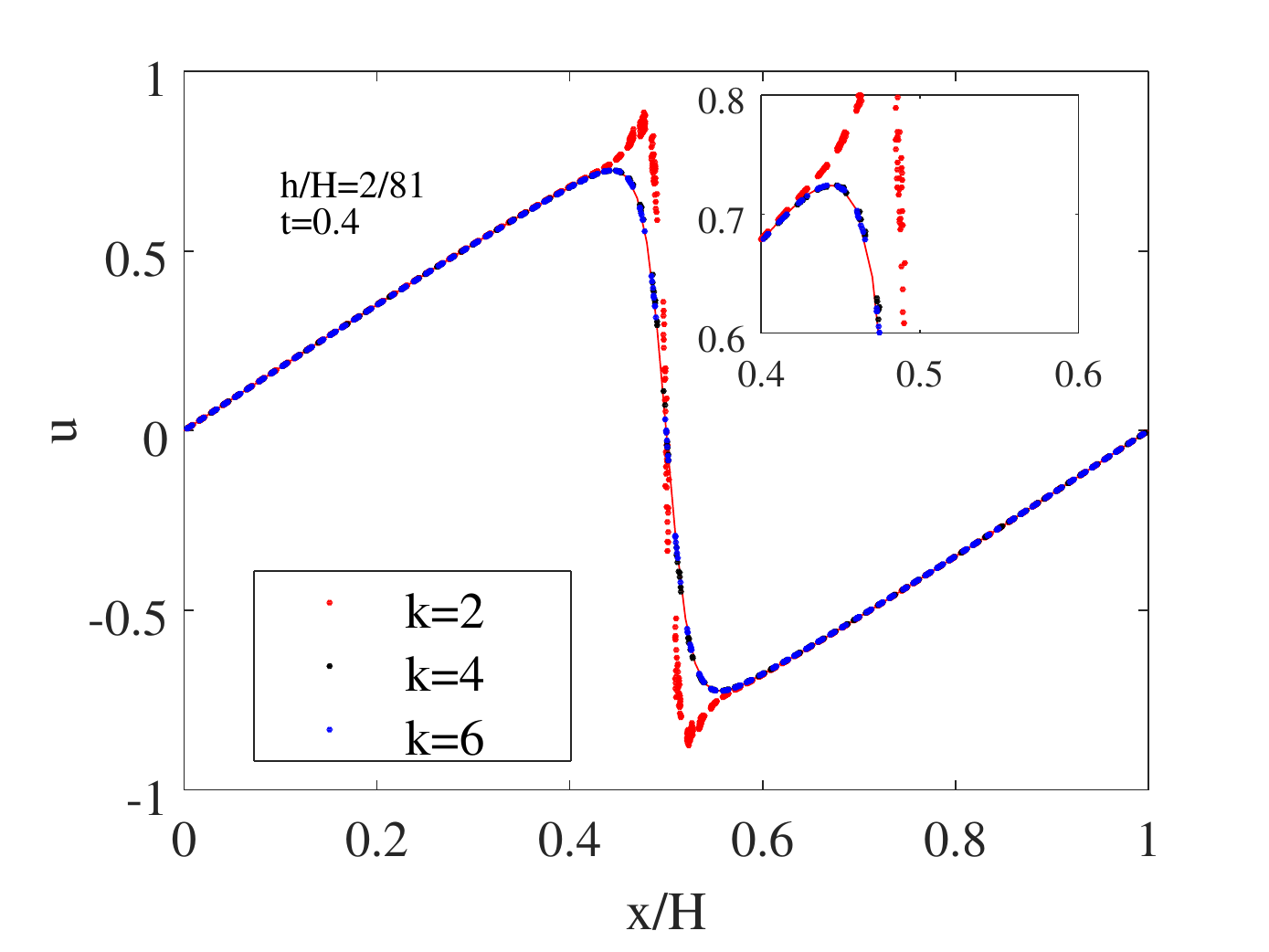}
\caption{The analytic (red line) and numerical solutions (coloured dots) of the periodic Burgers' problem for a range of parameters, with $Re=100$, $h/\delta{r}=2$, and $\varepsilon/\delta{r}=0.5$. Left panel: the effects of varying resolution on the solution at $t=0.4$, with $k=4$. Right panel: The effects of varying $k$ on the solution at $t=0.4$, with $h/H=2/81$.\label{fig:burgers_sol}}
\end{figure}

Figure~\ref{fig:burgers_sol} (left panel) shows the solution at $t=0.4$ with $k=4$ as the resolution is varied. For the coarse resolution of $h/H=2/41$ (blue dots) there is significant noise in the solution, as the shock is under-resolved, although the simulation remains stable. The right panel of Figure~\ref{fig:burgers_sol} shows a single resolution $h/H=2/81$, as $k$ is increased. For $k=2$ (red dots) the solution contains overshoots around the shock. This is because the eigenvalues of the Laplacian operator with $k=2$ have smaller negative real part than those for higher $k$, and hence are less able to dissipate the high order modes present in the solution. For $k=4$ and $k=6$ we see a close match with the analytic solution (thin red line).   

\begin{figure}
\includegraphics[width=0.6\textwidth]{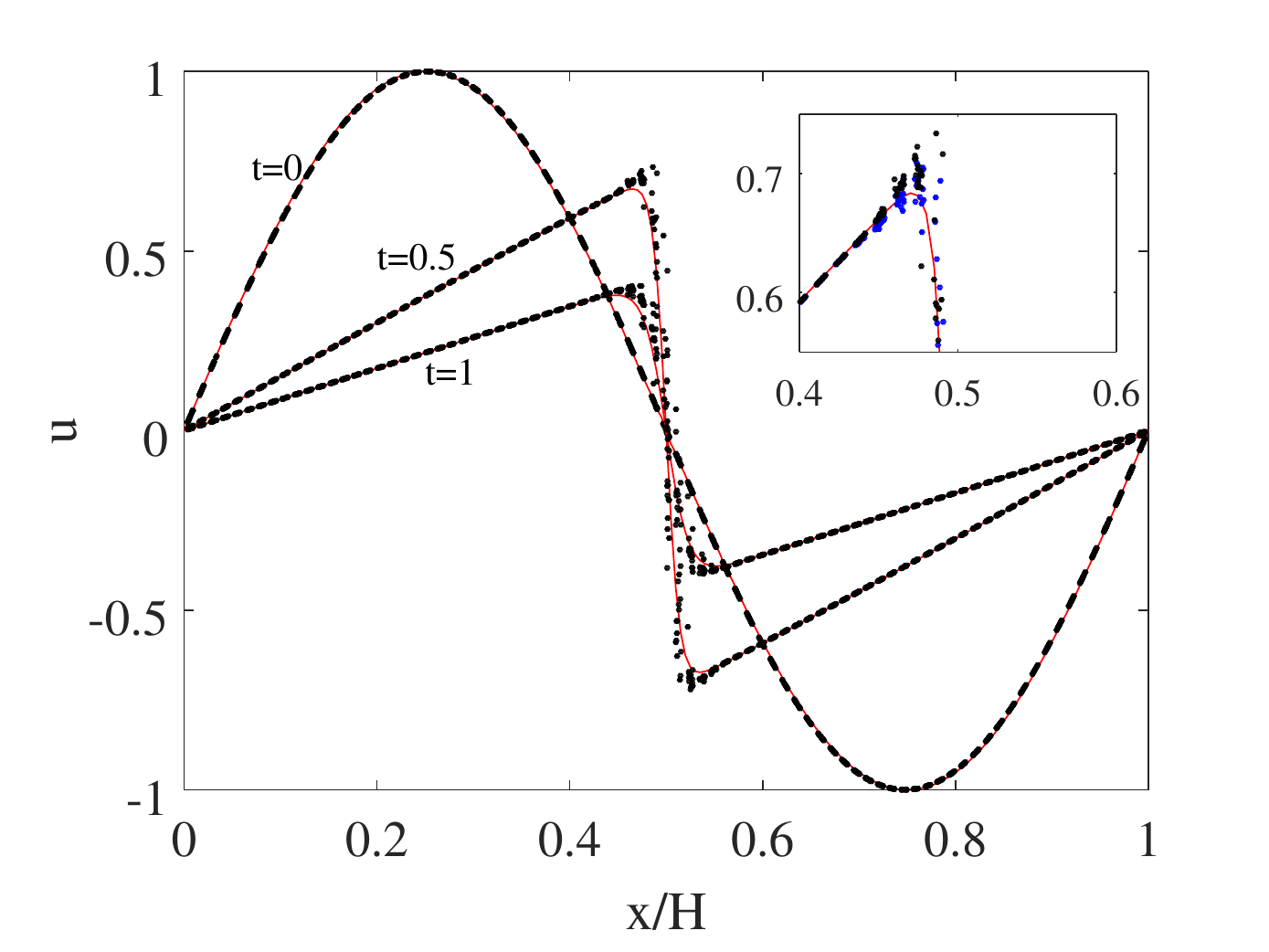}
\caption{The analytic (red line) and numerical solutions (coloured dots) of the periodic Burgers' problem with $Re=250$, $h/\delta{r}=2$, and $\varepsilon/\delta{r}=0.5$, and $k=6$. The solution is shownat times $t=0$, $t=0.5$, and $t=1$, obtained with a resolution of $h/H=2/81$. Black dots indicate biharmonic hyperviscosity, and blue dots (inset) indicate triharmonic hyperviscosity.\label{fig:burgers_sol_re250}}
\end{figure}

We increase the Reynolds number to $Re=250$, and again run the simulation with $\varepsilon/\delta{r}=0.5$, $h/\delta{r}=2$, $k=6$ and $h/H=2/81$. The simulation becomes unstable prior to $t=0.4$. By adding hyperviscosity to the scheme as described in Section~\ref{sec:stab}, we are able to stabilise the simulations, whilst retaining relatively good accuracy. Figure~\ref{fig:burgers_sol_re250} shows the solution at several times using a biharmonic hyperviscosity of $-100h^{4}\nabla^{4}$ (black dots). The inset shows detail of the solution at $t=0.5$ using the biharmonic hyperviscosity, and a triharmonic hyperviscosity of $8h^{6}\nabla^{6}$ (blue dots). The coefficients of hyperviscosity were chosen by numerical experiment as approximately the miminum values which stabilised the solution. Both forms of hyperviscosity are capable of stablising the solution, whilst retaining relatively good accuracy. In the present case we see only minor differences between biharmonic and triharmonic hyperviscosity. 

\begin{figure}[h]
\includegraphics[width=0.52\textwidth]{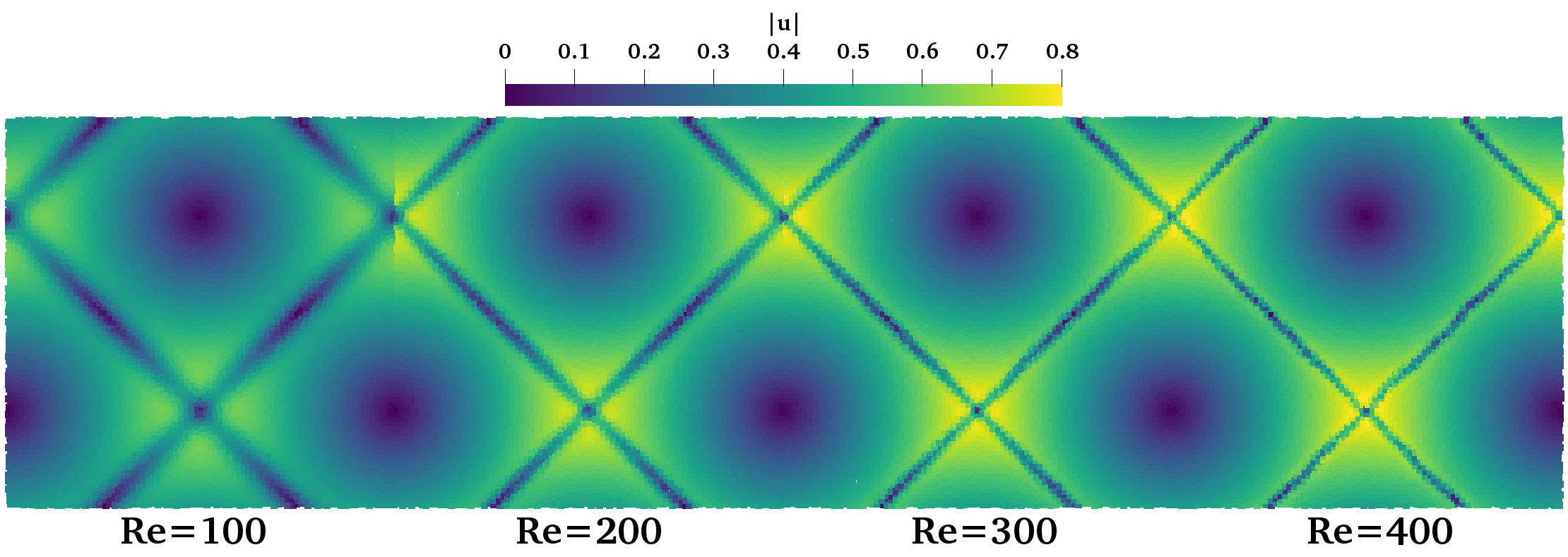}
\caption{The velocity magnitude field of the periodic Burgers' problem with two-dimensional initial conditions at $t=0.4$ for a range of $Re$. The solution is obtained with $h/H=2/41$, $\varepsilon/\delta{r}=0.5$, $h/\delta{r}=2$ and $k=6$.\label{fig:b_sinsin}}
\end{figure}

We now modify our initial conditions to the two-dimensional sinusoidal functions
\begin{equation}u=\sin\left(\frac{2\pi{x}}{H}\right)\sin\left(\frac{2\pi{y}}{H}\right);\quad{v}=-\cos\left(\frac{2\pi{x}}{H}\right)\cos\left(\frac{2\pi{y}}{H}\right)\end{equation}  
and run the simulation for a range of Reynolds numbers, up to $Re=400$. We do not have an analytic solution for the two-dimensional initial conditions, although intuitively it can be seen that certain slices through the domain must have an analytic solution which matches (or is a translation of) the solution for the one-dimensional initial conditions. Figure~\ref{fig:b_sinsin} shows the solution at $t=0.4$ for several values of $Re$, with a resolution of $h/H=2/41$ corresponding to $6561$ nodes in the domain. In cases, we set $k=6$, $h/\delta{r}=2$, and $\varepsilon/\delta{r}=0.5$. In ordinary use, we found a limit of $Re\approx150$ at this resolution (the limit increased with increasing resolution), due to the lack of upwinding in the LABFM. With biharmonic hyperviscosity the simulations are stable up to $Re=400$. For this Reynolds number (rightmost panel of Figure~\ref{fig:b_sinsin}), the width of the shock is approximately $h$, and despite this under-resolution, hyperviscosity provides stability without significant smoothing of the solution. 

\subsection{Navier-Stokes equations}

\begin{figure}
\includegraphics[width=0.8\textwidth]{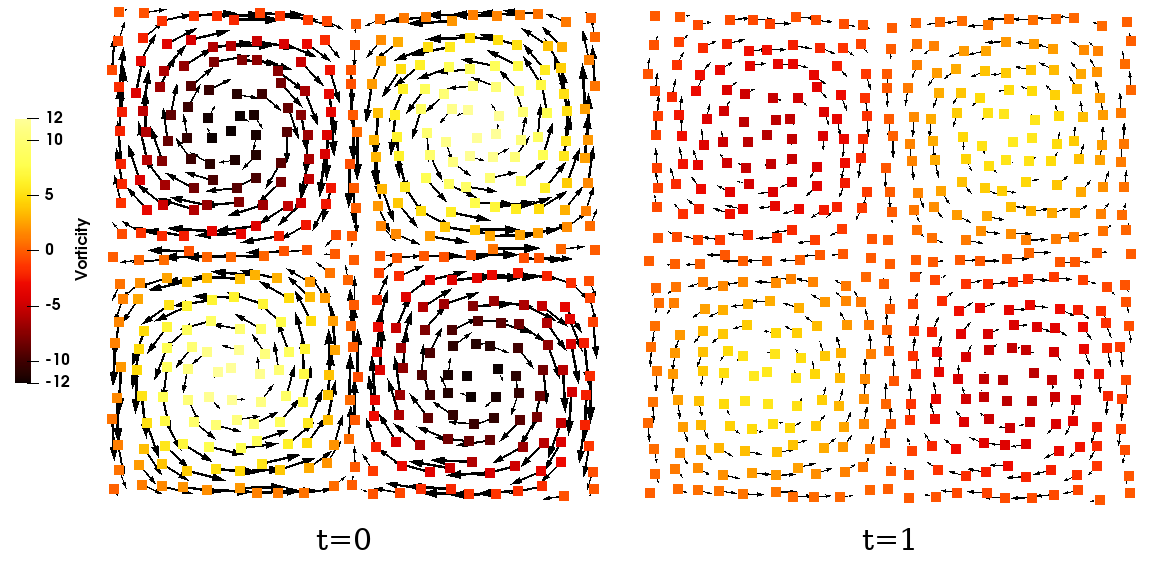}
\caption{The numerical solution to Taylor-Green vortices at times $t=0$ (left) and $t=1$ (right) for $Re=100$, on a node distribution with $h/H=2/11$ and $\varepsilon/\delta{r}=0.5$. Colour indicates vorticity, and arrows indicate velocity.\label{fig:tg_sol}}
\end{figure}

\begin{figure}
\includegraphics[width=0.7\textwidth]{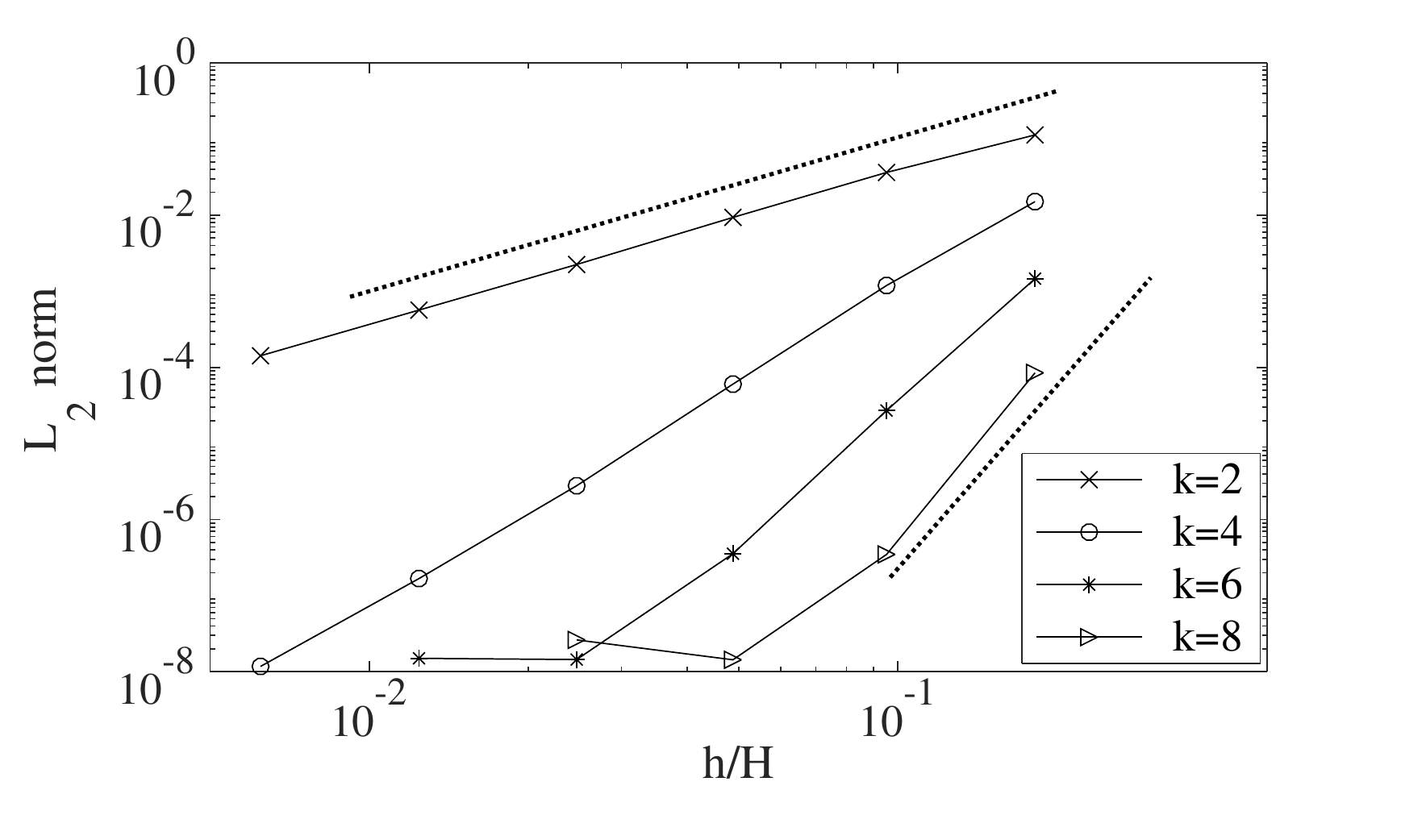}
\caption{Variation of the $L_{2}$-norm with resolution for the Taylor-Green vortices problem at time $t=1$, for different values of $k$. For $k=\left\{2,4,6\right\}$ we use $h/\delta{r}=2$ and $\varepsilon/\delta{r}=0.5$. For $k=8$ we use $h/\delta{r}=2.5$ and $\varepsilon/\delta{r}=0.2$. Simulations were run with $Re=100$, and using quadratic ABFs. The dotted lines show convergence rates of $2$ and $8$.\label{fig:ns_tg_conv}}
\end{figure}

Finally, we test the ability of our method to solve the incompressible Navier-Stokes equations. This test incorporates aspects of the previous PDEs, with the momentum equation being a non-linear hyperbolic-parabolic equation, and the incompressibility constraint being manifest as an elliptic PDE. Expressed in terms of the vorticity, the Navier-Stokes equations may be written
\begin{equation}\frac{\partial\omega}{\partial{t}}+\bm{u}\cdot\nabla\omega=\frac{1}{Re}\nabla^{2}\omega\label{eq:vort}\end{equation}
where $\omega$ is the vorticity and $\bm{u}$ is the velocity. Incompressibility is enforced by obtaining the velocity from a stream-function $\psi$, such that $\bm{u}=\nabla\times\psi$. The stream-function is obtained from the vorticity by solution of a Poisson's equation:
\begin{equation}\nabla^{2}\psi=-\omega\label{eq:vortpe}.\end{equation}
We integrate~\eqref{eq:vort} using RK4, and after each substep, solve~\eqref{eq:vortpe} to obtain $\psi$, then calculate gradients of $\psi$ to obtain $\bm{u}$ for use in the next substep. We set the time step using~\eqref{eq:tstep}. In the present study, we consider the Navier-Stokes equations in unbounded domains. The vorticity-streamfunction formulation is chosen for its ease of implementation in high-order explicit time-stepping schemes. The case we study here is that of Taylor-Green vortices. The initial conditions and solution are given by
\begin{equation}\omega\left(x,y,t\right)=\frac{4\pi}{H}\sin\left(2\pi{x}/H\right)\sin\left(2\pi{y}/H\right)\exp\left(\frac{-8\pi^{2}t}{Re{H}^{2}}\right).\end{equation} 
The velocity field is orthogonal to the gradient of the vorticity field, and so the advection term in~\eqref{eq:vort} is zero for the analytic solution, and the vorticity field decays exponentially. Figure~\ref{fig:tg_sol} shows the solution at times $t=0$ and $t=1$. For $Re=100$, the maximum vorticity decays from $4\pi$ (approximately $12.6$) to $5.7$. Figure~\ref{fig:ns_tg_conv} shows the variation of the $L_{2}$-norm with resolution at time $t=1$ with $Re=100$, for different values of $k$.  The coarsest resolution shown in Figure~\ref{fig:ns_tg_conv} is twice as coarse as the results shown in Figure~\ref{fig:tg_sol}. For $k=\left\{2,4,6\right\}$ we use $h/\delta{r}=2$ and $\varepsilon/\delta{r}=0.5$, whilst for $k=8$ we use $h/\delta{r}=2.5$ and $\varepsilon/\delta{r}=0.2$ (for stability). The errors converge with approximately order $k$, as expected. The clear error limit of approximately $10^{-8}$ is due to the accumulation of errors in the time stepping scheme. The dominant errors contributing to this limit are those originating in the approximate solution of~\eqref{eq:vortpe}, where the tolerance of the linear solver is set to $10^{-10}$. We find the rates of convergence remain when the Reynolds number is reduced to $Re=10$ or increased to $Re=1000$.

Whilst in the present study we have considered the Navier-Stokes equations only in unbounded domains, these results clearly demonstrate the potential of LABFM as a viable mesh-free method for computational fluid dynamics simulations.
In complex geometries the vorticity-stream function formulation chosen here requires careful (and computationally expensive) boundary treatment~\cite{rempfer_2006}, and for three-dimensional flows becomes more expensive still, as a Poisson equation must be solved for each component of the stream function vector field. Due to the collocated nature of this method, a primitive variable formulation is unstable, as the LBB condition~\cite{guermond_1998} is not satisfied. The development of a robust primitive variable formulation for LABFM is an ongoing area of research: a further study is planned involving the development of the LABFM framework for the solution of the Navier-Stokes equations in bounded complex geometries, with adaptivity in resolution and accuracy. The pairwise symmetry which leads to the conservation properties of SPH is necessarily broken ($w^{d}_{ji}\ne{w}_{ij}^{d}$) in LABFM, as the node distribution is accounted for to yield consistency. Hence, the method does not exhibit exact conservation: global conservation errors are determined by the order of the spatial discretisations $k$, the time integration scheme, and any boundary conditions. Based on finite volume concepts, a number of authors have developed consistent mesh-free methods which also exhibit exact conservation~\cite{chiu_2012,hopkins_2015,trask_2020}), though these have been limited to first or second order convergence rates. Such methods have significant potential for the accurate, efficient and robust simulation of flows in complex geometries, and an implementation of LABFM such that both exact conservation and high order consistency are retained is the subject of ongoing research.

\section{Conclusions}\label{conc}

In this work we have presented a novel framework for the calculation of high order finite differences on disordered node distributions. We have shown the ability of our method, referred to as the Local Anisotropic Basis Function method (LABFM), to generate up to eighth order discrete operators with compact computational stencils, which we have used to solve a range of prototype PDEs. Our method has several attractive characteristics:
\begin{enumerate}
\item The accuracy of the approximations holds even for node distributions with significant local anisotropy, provided the ABFs contributing to the discrete operator are adequately sampled over the computational stencil. In practice, in two dimensions we can obtain $4^{th}$ order with $\mathcal{N}\approx25$, and $6^{th}$ order with $\mathcal{N}\approx48$, and $8^{th}$-order with $\mathcal{N}\approx60$ on severely disordered distributions. This characteristic will allow the method to remain extremely accurate even in complex geometries where non-uniformity in the node distribution is unavoidable. In this regard the method has similarities with RBF-FD, although a benefit of LABFM is our ability to exactly prescribe the order or accuracy level. Furthermore, in the limiting low order case, our method collapses to SPH. Hence there is significant scope for a naturally adaptive (in resolution, accuracy and frame of reference) ALE scheme.
\item In LABFM we solve a linear system to obtain weights for a discrete operator, rather than to directly obtain derivative approximations as in FPM and CSPM. Therefore for Eulerian schemes the method is extremely fast, matching the $\mathcal{O}\left(N\right)$ FLOPs per time-step of RBF-FD, and for fourth order and higher, allowing smaller $\mathcal{N}$. In certain conditions LABFM is equivalent (in terms of errors) to certain forms of CSPM, but the generalisation to an arbitrary set of ABFs yields increased stability and resilience to node disorder as the order of the scheme is increased. 
\item As with RBF-FD, one-sided discrete operators are automatically generated at truncated domain boundaries, with the same order as the internal scheme, provided the node distribution adequately samples the set of ABFs used. Hyperviscosity, which effectively filters out unphysical short wavelength disturbances, can also be easily included in the discrete operators in LABFM, providing a simple means to increase stability for hyperbolic problems.
\item The notation describing the construction of LABFM is concise and easy to follow, and the implementation of the method into existing mesh-free codes should be straightforward.
\end{enumerate}
These characteristics give the method significant potential to be a powerful tool for the solution of PDEs in complex geometries. By virtue of its local nature, we expect LABFM to be highly scalable and computationally efficient, properties which will be explored in a further study. Whilst we have demonstrated up to eighth order schemes in the present study, this need not be the limit, provided computational stencils are increased as required. A further study, applying LABFM to solve the Navier Stokes equations in complex geometries is ongoing, and future work will involve the extension of the framework to include adaptivity in resolution and accuracy.

\section*{Acknowledgements}
We are grateful for financial support from the Leverhulme Trust, via Research Project Grants RPG-2017-144 and RPG-2019-206, and EPSRC via grant EP/R005729/1. We thank Georgios Fourtakas for several useful discussions regarding node distributions. We would like to thank two anonymous reviewers for very helpful comments, both regarding the current manuscript and our future work.

\appendix
\section{Details of ABFs used\label{rbf_details}}

Here we provide details of the ABFs investigated in this study, which are based on the following fundamental RBFs: conic, quadratic, Wendland C6, and Gaussian. Complete details of the ABFs based on the conic RBF are provided in the main body of the paper, in Section~\ref{abf}. All fundamental RBFs except the Gaussian have been normalised to yield $W^{0}\left(2\right)=0$, and
\begin{equation}2\pi\displaystyle\int_{0}^{2}qW^{0}\left(q\right)dq=1\label{eq:unity}.\end{equation}
The series of ABFs generated from $W^{0}$ are given by
\begin{multline}\bm{W}_{ji}=\frac{dW^{0}}{dr}\left[\frac{x}{r},\frac{y}{r},\frac{y^{2}}{r^{3}},\frac{-xy}{r^{3}},\frac{x^{2}}{r^{3}},\frac{-3xy^{2}}{r^{5}},\frac{2x^{2}y-y^{3}}{r^{5}},\frac{2xy^{2}-x^{3}}{r^{5}},\right.\\
\left.\frac{-3x^{2}y}{r^{5}},\frac{12x^{2}y^{2}-3y^{4}}{r^{7}},\frac{9xy^{3}-6x^{3}y}{r^{7}},\frac{2x^{4}-11x^{2}y^{2}+2y^{4}}{r^{7}},\frac{9x^{3}y-6xy^{3}}{r^{7}},\frac{12x^{2}y^{2}-3x^{4}}{r^{7}}\dots\right]^{T} + \\
\frac{d^{2}W^{0}}{dr^{2}}\left[0,0,\frac{x^{2}}{r^{2}},\frac{xy}{r^{2}},\frac{y^{2}}{r^{2}},\frac{3xy^{2}}{r^{4}},\frac{y^{3}-2x^{2}y}{r^{4}},\frac{x^{3}-2xy^{2}}{r^{4}},\frac{3x^{2}y}{r^{4}},\dots\right]^{T} + \\
\frac{d^{3}W^{0}}{dr^{3}}\left[0,0,0,0,0,\frac{x^{3}}{r^{3}},\frac{x^{2}y}{r^{3}},\frac{xy^{2}}{r^{3}},\frac{y^{3}}{r^{3}},\dots\right]^{T}+\frac{d^{4}W^{0}}{dr^{4}}\left[\dots\right]^{T}+\dots\end{multline}
In practice we omit the terms in $d^{2}W^{0}/dr^{2}$ and higher derivatives for all $k\ge4$ (i.e. for elements $10$ and onwards of $\bm{W}_{ji}$), regardless of whether $d^{2}W^{0}/dr^{2}$ and higher derivatives are zero. We do this because: a) it is computationally simpler, b) it yields a scheme which is more resilient to node distribution disorder. 

\subsection{Convex Quadratic}

The fundamental RBF is set as $W^{0}=W^{Q}$, where
\begin{equation}W^{Q}\left(q\right)=\frac{3}{16\pi}\left(q-2\right)^{2}\qquad0\le{q}\le2.\end{equation}
The design of $W^{Q}$ is such that $\left.dW^{Q}/dq\right\rvert_{q=2}=0$, which appears from our stability analysis to be a necessary condition for stability of the convective operator on a uniform node distribution. $W^{Q}$ is convex, with positive curvature everywhere and discontinuous gradient (i.e. a sharp peak) at the origin. Note that $W^{Q}$ is only valid for $q\in\left[0,2\right]$.

\subsection{Wendland C6}

The fundamental RBF is set as $W^{0}=W^{W}$, where 
\begin{equation}W^{W}\left(q\right)=\frac{78}{28\pi}\left(1-\frac{q}{2}\right)^{8}\left(4q^{3}+6.25q^{2}+4q+1\right)\qquad0\le{q}\le2,\end{equation}
which is the Wendland C6 kernel, (see~\citet{dehnen_aly}) commonly used in SPH. Note that $W^{W}$ is only valid for $q\in\left[0,2\right]$.

\subsection{Gaussian}

The fundamental RBF is set as $W^{0}=W^{G}$, where
\begin{equation}W^{G}\left(q\right)=\frac{9}{\pi}e^{-9q^{2}},\end{equation}
which is the classical Gaussian kernel, similar to that shown in~\cite{gingold_1977,lind_2016}. Note that for $W^{G}$, satisfaction of~\eqref{eq:unity} is not exact, but is approximate to within $2\times10^{-16}$, and $W^{G}\left(2\right)\approx6.6\times10^{-16}$.

\bibliographystyle{elsarticle-num-names}
\bibliography{jrckbib}

\end{document}